\documentclass{article}

\usepackage{arxiv}

\usepackage[utf8]{inputenc} 
\usepackage[T1]{fontenc}    
\usepackage{hyperref}       
\usepackage{url}            
\usepackage{booktabs}       
\usepackage{amsfonts}       
\usepackage{nicefrac}       
\usepackage{microtype}      
\usepackage{lipsum}		
\usepackage{graphicx}
\usepackage{natbib}
\usepackage{doi}

\usepackage{bbm}
\usepackage{amsmath}
\usepackage{makecell}
\usepackage{caption}
\usepackage{subcaption}

\usepackage{color}
\usepackage{colortbl}
\usepackage{multirow}

\newtheorem{theorem}{Theorem}
\newtheorem{lemma}{Lemma}
\newtheorem{remark}{Remark}

\newtheorem{definition}{Definition}
\newtheorem{corollary}{Corollary}
\newtheorem{simulation procedure}{Simulation procedure}

\title{CGR-CUSUM: A Continuous time Generalized Rapid Response Cumulative Sum chart}

\author{ \href{ https://orcid.org/0000-0001-9011-3743}{\includegraphics[scale=0.06]{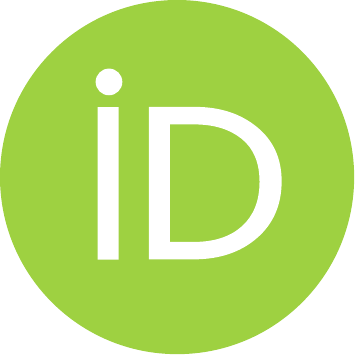}\hspace{1mm}Daniel Gomon}\thanks{To whom correspondence should be adressed.} \\
	Mathematical Institute\\
	Leiden University\\
	Leiden, the Netherlands \\
	\texttt{d.gomon@math.leidenuniv.nl} \\
	\And
	\href{ https://orcid.org/0000-0001-5395-1422}{\includegraphics[scale=0.06]{orcid.pdf}\hspace{1mm}Hein Putter} \\
	Department of Biomedical Data Sciences\\
	Leiden University Medical Centre\\
	Leiden, the Netherlands\\
	\And 
	\href{  https://orcid.org/0000-0003-1228-4162}{\includegraphics[scale=0.06]{orcid.pdf}\hspace{1mm}Rob G. H. H. Nelissen} \\
	Department of Orthopaedic Surgery\\
	Leiden University Medical Centre\\
	Leiden, the Netherlands\\
	\And 
	\href{ https://orcid.org/0000-0002-2448-5378}{\includegraphics[scale=0.06]{orcid.pdf}\hspace{1mm}Stéphanie van der Pas} \\
	Department of Epidemiology and Data Science\\
	Amsterdam UMC, Vrije Universiteit Amsterdam\\
	Amsterdam, the Netherlands\\
}

\renewcommand{\shorttitle}{\textit{arXiv} Template}

\hypersetup{
pdftitle={A template for the arxiv style},
pdfsubject={q-bio.NC, q-bio.QM},
pdfauthor={David S.~Hippocampus, Elias D.~Striatum},
pdfkeywords={CUSUM, Control charts, Survival analysis, Continuous time, Generalized likelihood ratio, Quality of care, Benchmarking},
}

\begin{document}
\maketitle

\begin{abstract}
Rapidly detecting problems in the quality of care is of utmost importance for the well-being of patients. Without proper inspection schemes, such problems can go undetected for years. Cumulative sum (CUSUM) charts have proven to be useful for quality control, yet available methodology for survival outcomes is limited. The few available continuous time inspection charts usually require the researcher to specify an expected increase in the failure rate in advance, thereby requiring prior knowledge about the problem at hand.  Misspecifying parameters can lead to false positive alerts and large detection delays. To solve this problem, we take a more general approach to derive the new Continuous time Generalized Rapid response CUSUM (CGR-CUSUM) chart. We find an expression for the approximate average run length (average time to detection) and illustrate the possible gain in detection speed by using the CGR-CUSUM over other commonly used monitoring schemes on a real life data set from the Dutch Arthroplasty Register as well as in simulation studies. Besides the inspection of medical procedures, the CGR-CUSUM can also be used for other real time inspection schemes such as industrial production lines and quality control of services.
\end{abstract}


\section{Introduction}
\label{sec1}

Rapid detection of deterioration in the quality of care can spare patients unnecessary health burdens.  There are currently many inspection schemes which can be used to monitor the quality of care, such as funnel plots \citep{spiegelhalter} and a variety of Cumulative Sum (CUSUM) charts (\citealp{cusumsteiner}; \citealp{biswas}). A particularly attractive property of CUSUM charts is that they can be used to sequentially check for a decrease in the quality of a process. Ideally, the inspection scheme is also tailored to the outcome type. In this article we are interested in inspecting survival outcomes, where every individual can experience a failure at any time after their entry into the study. As an example, the Dutch Arthroplasty Register (LROI) is interested in simultaneously monitoring the quality of orthopaedic care at multiple hospitals performing total hip replacement surgery by considering the information provided by the time of implant failure as soon as it occurs, as well as the information provided by patients not experiencing implant failures. To facilitate such real-time inspection, \citet{biswas} developed a CUSUM chart for survival outcomes, followed by \citet{RASTCUSUM} and \citet{NJR}. Each of these charts uses different assumptions in the CUSUM model and is therefore applicable in different scenarios. One similarity is that all of them require the researcher to specify an expected increase in the future rate of failure. When this quantity is chosen incorrectly, the charts may experience delays in detection and produce false negative signals.

Our main goal in this article is to develop a method that no longer requires the researcher to specify many parameters in advance, thereby requiring less prior information for inspection and leading to faster detection times in practical applications.  For this reason, we devise a generalisation of the CUSUM chart by \citet{biswas}, which we call the Continuous time Generalised Rapid Response CUSUM (CGR-CUSUM). \citet{biswas} chose to only consider the information provided by patients until $1$ year after their procedure. In contrast, the CGR-CUSUM is constructed using all available information on any patient at all times. A consequence of these changes is that generally our chart leads to quicker detection of underperforming hospitals, thereby contributing  to the improvement of the quality of care.

Other methods for the continuous time inspection of the quality of care include the uEWMA chart for survival time data by \citet{uEWMA} and the STRAND chart by \citet{STRAND}. \citet{STRAND} briefly discusses the differences between the BK-CUSUM, uEWMA and STRAND charts, and concludes that the uEWMA and STRAND charts are particularly suitable for quick detection when failures are clustered. In contrast, the BK-CUSUM and the CGR-CUSUM are designed to detect increased failure rates without a specific mechanism for clusters.

We derive an approximation for the average run length (average time to detection) of the CGR-CUSUM, by means of considering a simplification of the CGR-CUSUM called the Continuous time Generalised Initial response CUSUM (CGI-CUSUM). Additionally, we consider an adjusted \citet{biswas} CUSUM procedure which uses the information of all patients at all times, which we  call the BK-CUSUM for convenience. Similarly, we present an approximation to the average run length of the BK-CUSUM and compare this approximation with the approximation found for the CGR-CUSUM. This comparison demonstrates how incorrect prior information can significantly increase the detection times of the BK-CUSUM procedure, which then also carries over to the \citet{biswas} CUSUM.

The new CGR-CUSUM chart can be a very useful tool in practical applications where the future expected rate of failure is not known in advance or likely to vary over the time of the study.  As this occurs often in medical applications, the CGR-CUSUM chart can significantly improve the quality of care worldwide by inspecting current procedures. In contrast to the multi-chart CUSUM scheme of \citet{han_tsung}, where the possible increase in failure rate is considered over a finite probable domain, the CGR-CUSUM only requires the construction of one chart and the increase in failure rate can also be limited to a fixed domain. On top of this, the CGR-CUSUM is not limited to medical applications. The chart can be used to inspect any procedures involving ``survival" outcomes, such as production lines and customer satisfaction inspection.

 In Section \ref{sec2} of this article the relevant quantities, notation as well as the CGR-CUSUM are introduced. Afterwards the \citet{biswas} CUSUM procedure is introduced, as well as an adjusted version thereof which we call the BK-CUSUM. An approximate average run length is derived for both procedures.  In Section \ref{sec4} all methods are applied to a data set from the Dutch Arthroplasty Register (LROI). In Section \ref{sec3} a simulation study is performed to compare the average run lengths of aforementioned procedures under restrictions on their null (hypothesis) average run length. Additionally, a simulation study is performed using the  data from this register where the type I error of the charts over time is restricted under the null rate. The article concludes with a discussion and recommendations for practice.
 
 \vspace{-1.5em}

\section{Methods}
\label{sec2}

\subsection{Model and data}
Following \citet{biswas}, consider a hospital with subjects $i = 1, 2, ...$ arriving (entering the study) according to a Poisson process with rate $\psi$. Let $S_i$ denote the time of entry of subject $i$ into the study, relative to the starting time $t=0$. Denote by $X_i$ the time from entry until failure, such that $T_i = S_i + X_i$ is the chronological time of failure. Consider only right-censored observations, and let $R_i$ denote the chronological time of right-censoring of observation $i$.  Let the $p-$vector $\mathbf{Z}_i$ denote the relevant covariates of subject $i$. Assume that there is a known null-distribution for the subject specific time to failure, denoted by the hazard rate $h_i(x)$. 
We make use of the Cox proportional hazards model to incorporate the covariates, such that $h_i(x) = h_0(x) \exp{\left( \mathbf{Z}_i^\top \pmb{\beta} \right)}$ with regression coefficients $\pmb{\beta}$ and known baseline hazard rate $h_0$. Let $Y_i(t) = \mathbbm{1} \{ S_i \leq t \leq  \min\{T_i, R_i\} \}$ be an indicator whether subject $i$ is \emph{active} at time $t$. Define $\tilde{N}_i(t)= \mathbbm{1}\{T_i \leq t \}$ and subsequently define $N_i(t) = \int_0^t Y_i(u) d\tilde{N}_i(u)$ for $t > 0$ as the counting process for an observed failure of subject $i$. Define $N(t) = \sum_{i \geq 1} N_i(t)$ as the counting process for the total number of failures observed at the hospital. Define the cumulative intensity of subject $i$ as $\Lambda_i(t) = \int_0^t \lambda_i(u) du$ with $\lambda_i(u) = Y_i(u) \cdot h_i(u)$. Let the superscript $\theta$ indicate an increase in the hazard rate such that $\Lambda_i^\theta(t) = \Lambda_i(t) \cdot \exp(\theta)$ and $h_i^\theta(t) = h_i(t) \cdot \exp(\theta)$ and $F_i^\theta$ the associated cumulative distribution function.  We call $\exp(\theta)$ the \emph{hazard ratio} and say that the process is \emph{in control} when $\theta = 0$ and \emph{out of control} when $\theta > 0$. Define $\Lambda(t) = \sum_{i \geq 1} \Lambda_i(t)$ as the total cumulative intensity at the hospital at time $t$. For aforementioned counting processes, define $dN(t)= N(t + \mathrm{d}t) - N(t)$, with $\mathrm{d}t$ an infinitesimally small quantity. It follows that:
\begin{align}
    \mathbbm{P} \left( \mathrm{d}N_i^D(t) = 1| T_i \geq t, S_i, Z_i \right) = \begin{cases}
    e^\theta h_i(t-S_i) \mathrm{d}t, & \text{if } 0 \leq t -S_i, \\
    0, & \text{else.}     \label{eq:dN(t)}
    \end{cases}
\end{align} We denote the right-hand side of \eqref{eq:dN(t)} by $\mathrm{d}\Lambda_i^\theta(t)$.

\subsection{Continuous Time Generalized Rapid Response CUSUM (CGR-CUSUM)}

The CUSUM procedure developed by \citet{biswas} can be used to test whether the hazard rate at a hospital has increased from $\Lambda_i$ to  $\Lambda_i^\theta$ for some fixed and known $\theta > 0$, at some unknown time after the start of the study. This procedure is very useful when there is some prior knowledge about the true hazard ratio $e^\theta$, but may lead to delays in detection when this is not the case or when the rate of failure is variable. For this reason we will consider a more general test, where the expected hazard ratio no longer needs to be specified in advance, much like the GLR Statistic in \citet{Siegmund_1995} is a generalization of the original CUSUM procedure of \citet{page}. 

To achieve this, we test the hypotheses of a sudden change in hazard rate at some unknown study time $s > 0$, affecting all subjects at risk at time $s$ and thereafter:
\begin{equation}
\label{eq:BKhypotheses}
\begin{aligned}
H_0: X_i \sim \Lambda_i(t), i = 1,2,...  & &H_1: 
\begin{array}{l}
X_i \sim \left. \Lambda_i(t) \right| t < s , i = 1, 2, ... \\
X_i \sim \left. \Lambda_i^\theta(t) \right| t \geq s, i = 1, 2, ... 
\end{array}
\end{aligned}
\end{equation} with $\theta > 0$. Let us find the likelihood ratio for a test of $\theta = 0$ against $\theta = \theta_1$ with $\theta_1 > 0$ an unknown constant. The distribution of $\mathrm{d}N_i(t)$ is approximately Bernoulli distributed with probability  $\lambda_i^\theta(t) \mathrm{d}t$ conditional on the history up to time $t$ (e.g. \citet{klein}, Section 3.6). The likelihood of all $n$ observations is then equal to $\prod_{i =1}^n \left( \prod_{0 \leq u \leq t }  \lambda_i^\theta(u)^{\mathrm{d}N_i(u)} \right) \exp \left( -  \Lambda_i^\theta(t) \right)$ up to time $t$. This yields a likelihood ratio statistic at time $t$ of:
\begin{align*}
    \mathrm{LR}(t) &= \sup_{\theta \geq 0} \frac{\prod_{i =1}^n  \left( \prod_{0 \leq u \leq t } \exp(\theta) \lambda_i(u)^{\mathrm{d}N_i(u)} \right) \exp \left( -  \exp(\theta) \Lambda_i(t) \right)}{\prod_{i =1}^n \left( \prod_{0 \leq u \leq t } \lambda_i(u)^{\mathrm{d}N_i(u)} \right) \exp \left( -  \Lambda_i(t) \right)} \\
    &= \prod_{i=1}^n \frac{\left(\exp(\hat{\theta}(t)) \right)^{N_i(t)} \exp \left( -\exp(\hat{\theta}(t)) \Lambda_i(t) \right)}{\exp(-\Lambda_i(t))},
\end{align*} where $\hat{\theta}(t)$ is the maximum likelihood estimate of $\theta$ at time $t$. This maximum likelihood estimator $\hat{\theta}(t)$ can be determined by maximising the likelihood at a hospital where patients are failing with cumulative intensity $e^\theta \Lambda_i(t)$ up to time $t$ over $\theta$ and is given by:
\begin{align}
\label{thetatdef}
    \hat{\theta}(t) = \max \left(0, \log \left( \frac{N(t)}{ \Lambda(t)} \right) \right).
\end{align} The logarithm of the LR statistic is then given by:
\begin{align*}
    U(t):= \log(\mathrm{LR}(t)) &=  \hat{\theta}(t) N(t) - \left(\exp(\hat{\theta}(t)) -1\right)  \Lambda(t).
\end{align*}
 Note that this quantity will increase as soon as a failure is observed, and drift downwards at all other times.    A preliminary chart is then found by maximising $U(t)$ over the interval $[s,t]$:
\begin{align}
\label{eq:G(t)}
    G(t) &= \max_{0 \leq s \leq t} \left\lbrace \hat{\theta}(s,t) N\left( s,t \right) - \left( \exp\left(\hat{\theta}(s,t) \right) - 1 \right)  \Lambda \left( s, t\right) \right\rbrace,
\end{align} where $s$ indicates that the quantity is determined using the information provided by all active patients in the time frame $(s, t)$:
\begin{align}
\label{notG(t)}
\begin{array}{l}
N(s,t) = N(t) - N(s) \\
\hspace{0.3em} \Lambda(s,t) = \Lambda(t) - \Lambda(s)
\end{array} \text{ and } \text{ } 
\hat{\theta}(s,t) = \max \left(0, \log \left( \frac{N(s,t)}{ \Lambda(s,t) } \right) \right). 
\end{align}In contrast to the method developed by \citet{biswas}, it is not straightforward to determine this chart recursively due to the combination of a maximum likelihood estimator over time and the maximisation term over recent time. This makes the chart very computationally expensive. We therefore consider simpler hypotheses:
\begin{equation}
\label{eq: hypotheses}
\begin{aligned}
H_0: X_i \sim \Lambda_i, i = 1,2,...  & &H_1: 
\begin{array}{l}
X_i \sim \Lambda_i, i = 1, 2, ..., \nu -1 \\
X_i \sim \Lambda_i^\theta, i = \nu, \nu +1, ... .
\end{array}
\end{aligned}
\end{equation}with $\theta > 0$ and $\nu \geq 1$ both unknown in advance.  We then test whether the rate of failure at the hospital has changed to $e^\theta \Lambda_i$, starting from some subject $\nu \geq 1$. These hypotheses make sense in a medical context, where the hazard rate is likely to depend on the entry time of the patient.
\begin{definition}
The Continuous time Generalised Rapid response CUSUM (CGR-CUSUM) chart is given by:
\begin{align}
        \mathrm{CGR}(t) &= \max_{1 \leq \nu \leq n} \left\lbrace \hat{\theta}_{\geq \nu}(t) N_{\geq \nu}(t) - \left( \exp\left(\hat{\theta}_{\geq \nu}(t)\right) - 1 \right)  \Lambda_{\geq \nu}(t) \right\rbrace
        \label{def:CGRCUSUM}
\end{align} with (subjects sorted according to chronological arrival time):
\begin{align}
\label{notCGR}
\begin{array}{l}
N_{\geq \nu}(t) = \sum_{i \geq \nu} N_i(t) \\
\hspace{0.3em} \Lambda_{\geq \nu}(t) = \sum_{i \geq \nu} \Lambda_i(t)
\end{array} \text{ and } \text{ }
\hat{\theta}_{\geq \nu}(t) &= \max \left(0, \log \left( \frac{N_{\geq \nu}(t)}{ \Lambda_{\geq \nu}(t) } \right) \right).
\end{align}
\end{definition}
 In the CGR-CUSUM patients prior to the $\nu-$th patient no longer contribute to the chart at all, whereas in $G(t)$ all patients active after time $t-s$ still contribute to the value of the chart. This difference is highlighted in Figure 1 of the Supplementary Materials. To employ a testing procedure, we construct the chart $\mathrm{CGR}(t)$ at every relevant time point $t$ and reject the null hypothesis (producing a signal) as soon as $\mathrm{CGR}(t) \geq h$ for some $h > 0$. This constant $h$ is called the \emph{control limit}, and can be chosen in accordance to some desired property of the procedure such as the average run length of the chart defined below.
\begin{definition}
Denote by $\tau_h = \inf\{t > 0: \mathrm{CGR}(t) \geq h\}$ the time it takes for a CGR-CUSUM to produce a signal. The average run length (ARL) is then defined as $\mathbbm{E}[\tau_h]$.  We refer to the \textit{in control} average run length as the expected time to detection when $\exp(\theta) = 1$ and \textit{out of control} average run length when $\exp(\theta) > 1$.
\end{definition}

\subsection{An approximation to the ARL}
\label{subsec:approxARL}

In this section we will derive an upper bound for the average run length of the CGR-CUSUM in the out of control case. The maximisation term in \eqref{def:CGRCUSUM} poses a significant challenge in approximating the ARL. It turns out that we can derive a bound on the ARL through comparison with a simpler version of the CGR chart. For this reason we consider the Continuous time Generalised Initial response (CGI) CUSUM chart. This chart can be used to test the hypotheses of an initial change in the rate of failure:
\begin{align*}
    H_0: &X_i \sim \Lambda_i, i = 1, 2, ... &H_1:& X_i \sim \Lambda_i^\theta, i = 1, 2, ... .
\end{align*} 
\begin{definition}
The Continuous time Generalized Initial response CUSUM (CGI-CUSUM) with $\hat{\theta}(t)$ as in \eqref{thetatdef} is given by:
\begin{align*}
    \mathrm{CGI}(t) &=  \hat{\theta}(t) N(t) - \left( \exp(\hat{\theta}(t)) - 1 \right) \Lambda(t).
\end{align*} 
\end{definition}
 Note how the CGI chart is simply the CGR chart without the maximisation term. The CGI-CUSUM is not a chart which should be used in practice as it cannot be used to sequentially detect a changepoint in the process, but instead it is merely a tool for theory. Due to its simpler expression, it is possible to determine the asymptotic distribution of the chart under some assumptions. One of the key assumptions is that subjects arrive according to a Poisson process with rate $\psi$, allowing us to equate the number of patients to time by $n \approx \psi \cdot t$. 
\begin{theorem}
Suppose that subjects arrive according to a Poisson process with rate $\psi$ under suitable regularity conditions. Then, for $\theta > 0$:
\begin{align*}
    \sqrt{t} \left( \mathrm{CGI}(t) - (\theta + \exp(-\theta) -1) I(\theta, t) \right) \overset{d}\rightarrow \mathcal{N} \left( 0, t \theta^2 I(\theta, t)  \right),
\end{align*} and when $\theta = 0$ (using the shape $k$/scale $b$ parametrisation):
\begin{align*}
    t \cdot \mathrm{CGI}(t) \overset{d}\rightarrow \Gamma \left(k = \frac{1}{2}, b = t \right)
\end{align*} where $ I(\theta, t) = \psi \int_0^t \mathbbm{E}_{Z_i} \left[  F_i^\theta(s) \right] ds$ is the Fisher information in all observations at time $t$.
\end{theorem}
 The proof of this theorem, the required regularity conditions as well as the derivation of the Fisher information can be found in the Supplementary materials Sections 2, 3 and 4. The usefulness of this result depends on the availability of an expression for $I(\theta, t)$. A discussion on how to calculate the Fisher information, as well as some examples for the PVF family of distributions can be found in the Supplementary Materials Section $7$. We determine an approximate (asymptotic) average run length for the CGI chart by equating the expected value of the asymptotic distribution to the control limit $h$. 

\begin{lemma}
We find an approximate average run length $\widehat{\mathrm{ARL}}_{\mathrm{CGI}}(\theta, h)$ for the CGI-CUSUM when $\exp(\theta) > 1$ by solving the following equation for $t$:
\begin{align}
    \left(\theta + \exp(-\theta) -1\right) I(\theta, t) = h.
    \label{eq:CGIARL}
\end{align}
\end{lemma}
For $\exp(\theta) = 1$, this method yields no approximation to the ARL and it is therefore not possible to determine theoretical control limits which restrict the in control ARL. It is possible to approximate the value of the in control average run length by means of Monte Carlo simulation when it is of interest. Note that due to the convergence requirement, this approximate ARL will not yield good approximations for small values of the control limit $h$. The theoretical out of control ARL will be evaluated by means of simulation in Section \ref{sec:comparARL}.

Note that the CGR-CUSUM is simply a CGI-CUSUM maximised over the last $n-\nu$ patients. As a result, the CGR-CUSUM is always larger or equal than the CGI-CUSUM. This property allows us to compare the average run lengths of the CGR- and CGI-CUSUM charts.
\begin{remark}
Suppose that subjects are failing with increased hazard rate $\Lambda_i \exp(\theta)$ from the beginning of the study. Then the  average run lengths of the charts can be compared as follows:
\begin{align*}
    \mathrm{ARL}_{\mathrm{CGR}}(\theta, h) \leq \mathrm{ARL}_{\mathrm{CGI}}(\theta, h).
\end{align*}
\end{remark}
 In most practical applications, an upper bound is sufficient as the interest lies in restricting the run time of the chart from above when the failure rate is higher than expected. 

Due to the found upper bound, we can now determine the CGI chart on out of control samples in simulation studies to obtain information on the ARL of the CGR chart for comparable samples. This negates the need to construct the CGR chart when approximating the ARL, saving a lot of computation time. Another way to reduce the computation time of the CGR- and CGI-CUSUM charts is given in the following corollary.

\begin{remark}
The value of the CGR-CUSUM and CGI-CUSUM can only increase at a time point when a failure is observed. As a consequence, for detection purposes it is sufficient to only determine the value of the charts at the times of failure.
\end{remark}

\subsection{The \citet{biswas} CUSUM and CGR-CUSUM}

By a priori fixing a value $\theta_1 > 0$ for $\theta$ in the chart $G(t)$ (see Equation \eqref{eq:G(t)}) we would recover the CUSUM procedure developed by \citet{biswas}. The biggest advantage of the CGR-CUSUM over the \citet{biswas} CUSUM is that we no longer need to specify this expected hazard ratio, allowing for a more general test requiring less prior knowledge. Besides this, the maximum likelihood estimator allows for updating the parameter to the most recent failure rates. In contrast, the maximum likelihood estimator needs time to converge to the true value, possibly causing delays in detection when compared to the \citet{biswas} CUSUM with correctly specified $\theta_1$.

\citet{biswas} note that one year post procedure survival outcomes are often employed for medical inspection schemes, and decide to consider subjects as active only for $C = 1$ year after the procedure. This limitation allows them to derive a closed form approximation to the average run length of the chart.  We decide not to disregard the information provided by patients $1$ year post procedure. The value of the chart is then based on more complete information, possibly leading to quicker detection times. With this approach, determining an expected run length shorter than $C = 1$ year is possible, in contrast to \citet{biswas}. Our new approach then also leads towards an approximate ARL for the \citet{biswas} CUSUM procedure with the $C=1$ limitation relaxed. Further on in this article, we will only consider the \citet{biswas} CUSUM procedure with the $C=1$ limitation relaxed, as it is more similar to our CGR chart. We call this chart the BK-CUSUM chart. 

\begin{definition}
The BK-CUSUM is given by:
\begin{align*}
    BK(t) &= \max_{0 \leq s \leq t} \left\lbrace \theta_1 N(s,t) - \left(  \exp(\theta_1) - 1 \right)  \Lambda(s,t) \right\rbrace,
\end{align*} with notation as in \eqref{notG(t)} where $\exp(\theta_1)$ is the expected hazard ratio chosen in advance.
\end{definition}

Taking a similar approach to Section \ref{subsec:approxARL}, it is possible to determine an approximate average run length for the BK-CUSUM procedure.
\begin{corollary}
Suppose $\theta_1$ is chosen such that $\exp(\theta_1)/\exp(\theta) < \theta_1 + \exp(-\theta)$. We find an approximate average run length $\widehat{\mathrm{ARL}}_{\mathrm{BK}}(\theta, h)$ by solving the following equation for $t$:
\begin{align}
    \left(\theta_1 + \exp(-\theta) - \frac{\exp(\theta_1)}{\exp(\theta)}\right) I(\theta, t) = h.
    \label{eq:BKARL}
\end{align}
\end{corollary}
The proof can be found in the Supplementary Materials section 5. 

 Due to the restriction on $\theta_1$ it is not always possible to use this expression for the approximate ARL. As $I(\theta, t)$ is non-negative for every $t \geq 0$, the approximate ARL for the CGR and the BK-CUSUM can be compared. It can easily be seen that when $\theta_1 \neq \theta$, the left side of \eqref{eq:CGIARL} is guaranteed to be larger than the left side of \eqref{eq:BKARL} for $t> 0$. This means that when the expected hazard ratio $\exp(\theta_1)$ is misspecified, the approximate ARL of the CGI chart will be smaller than that of the BK-CUSUM chart therefore yielding faster out of control detection speeds.

 The difference between the CGR-CUSUM and BK-CUSUM lies in the hypotheses used for constructing the chart, where the CGR-CUSUM is used to detect a change in hazard rate for all patients entering after some patient entry time and the BK-CUSUM to detect a spontaneous change in hazard rates for all patients at risk after some chronological time. This difference is shown visually in Figure 1 of the Supplementary materials.

\section{Application to Dutch Arthroplasty Register (LROI)}
\label{sec4}

We demonstrate the possible gain in detection speed when using the CGR-CUSUM over the BK-CUSUM by applying both methods on a hip replacement data set from the Dutch Arthroplasty Register (LROI). The LROI is the Dutch national registry of all orthopaedic implants (e.g. hip, elbow, wrist, ankle, knee, shoulder, finger, thumb), with a reported completeness of more than $95$ percent for registered hip and knee surgical procedures  (\citet{LROIcomplete}, \citet{lroisite}).

\vspace{-1.5em}

\subsection{The data set}
\label{subsec:dataset}
 The data used for the analysis consists of information on total hip replacement surgeries at $97$ hospitals across the Netherlands from $01/01/2014$ up until $01/01/2020$ and was received under agreement LROI $2020-053$. Available variables are the dates of all primary procedures, time until failure of the prosthesis (our main interest) and/or death of the patient as well as multiple characteristics of each patient which can be found in the Supplementary materials Table 2. Three characteristics of patients had more than $0.5$ percent of missing values, which were BMI $(1.8 \%)$, Smoking indicator ($4.5 \%)$ and Charnley Score $(5.3 \%)$. Using the R package mice (\citet{miceR}) we imputed missing values to obtain a complete data set.
 
 \vspace{-1em}
 
 \subsection{Baseline: yearly funnel plot}
 The current method employed by the LROI for comparing implant surgery performance between hospitals is a yearly risk-adjusted funnel plot over all available data of the recent three to six years. The funnel plot uses one year post surgery failure as binary outcome, therefore not allowing for continuous inspection of the quality of care. \citet{SchieCUSUM} have used the funnel plot as the ``golden standard'' for the LROI, indicating which hospitals had problems in their quality of care. As we have no information on the true failure rate and problems at the hospitals in question, we will compare detection times  with the funnel plot as well.
 
 
 \subsection{The baseline hazard}

In any practical application, the determination of the baseline is of great importance when considering the BK- and CGR-CUSUM charts as this greatly influences the detection speed and false detection rate. In both cases, the baseline is completely fixed by a null hazard rate and the corresponding Cox regression coefficients. In good scientific practice, these quantities should be determined using an in control data set, where failures are known to be happening at an acceptable rate. In reality, it is often difficult to obtain such a set for many different reasons. Because of this we determine the null hazard rate and Cox coefficients using the whole data set as training set. This implies that the average national failure rate over all hospitals is up to the desired standard. The same is done for the funnel plot.


The yearly funnel plot requires a yearly determination of the baseline. To make a fair comparison between the funnel plot and CUSUM methods, we therefore determine the baseline hazard rate, failure proportion and risk-adjustment coefficients using the whole data set restricted to the first $3$, $4$, $5$ and $6$ years of information for both methods. This was achieved by using the Cox proportional hazards function in the R package survival developed by \citet{survivalR}. This way, all methods use the same information for the construction of the charts and thereby all use the same ``standard of care".  We start with determining parameters at the $3$ year margin to have a sufficient amount of information for the determination of the null parameters.

\subsection{Determining control limits}

We determine control limits for the risk-adjusted BK- and CGR-CUSUM charts by restricting the simulated probability of a type I error to $0.05$ over a period of $6$ years. The procedure is described in the Supplementary materials Section $6$. Due to the extremely low failure rate in the data we chose to restrict $e^{\hat{\theta}(t)} \leq 6$. This comes down to believing the hazard rate at a hospital cannot be more than $6$ times the baseline. Without this limitation, the CGR-CUSUM made very large jumps for patients experiencing near instant failure (first or second day after surgery), almost always leading to detection. The determined control limits and a summary of detection times with respect to the hospitals detected by the funnel plot in the first three years can be found in Table \ref{table:detecttimessummary}. The continuous time procedures have their detection time rounded upwards to the closest month, to show what detection times are realistic when constructing the charts monthly.  We also include the detection times achieved by the monthly Bernoulli CUSUM procedure suggested by \citet{SchieCUSUM}. They chose to take a control limit of $h = 3.5$ in correspondence to other literature.
Exact detection times for all detected hospitals can be found in Table 1 of the Supplementary materials.

\subsection{Result: average detection delays}

\begin{figure}
\centering
 \begin{subfigure}[b]{0.48\textwidth}
        \includegraphics[width=\textwidth]{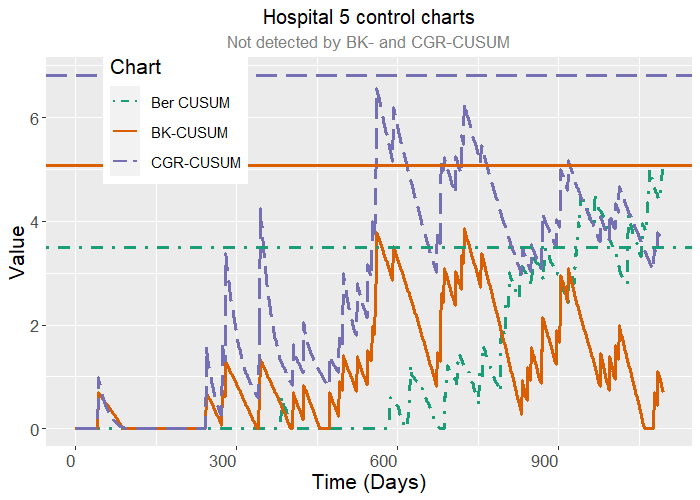}
        \caption{Hospital 5}
        \label{fig:Hosp5}
    \end{subfigure}
    \begin{subfigure}[b]{0.48\textwidth}
        \includegraphics[width=\textwidth]{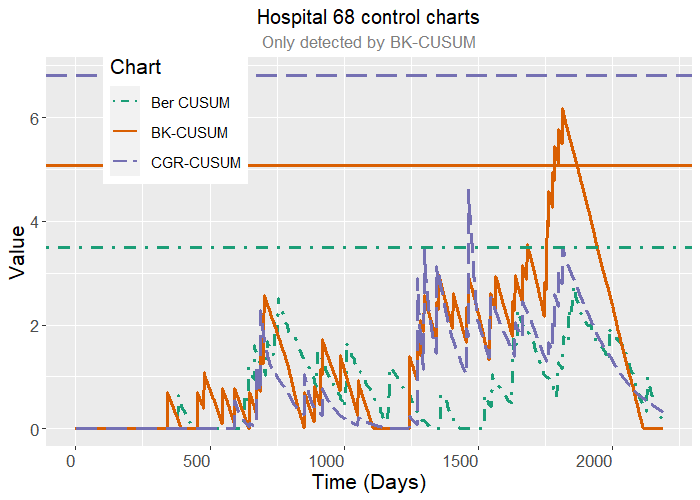}
        \caption{Hospital 68}
        \label{fig:Hosp68}
    \end{subfigure}
    \begin{subfigure}[b]{0.48\textwidth}
        \includegraphics[width=\textwidth]{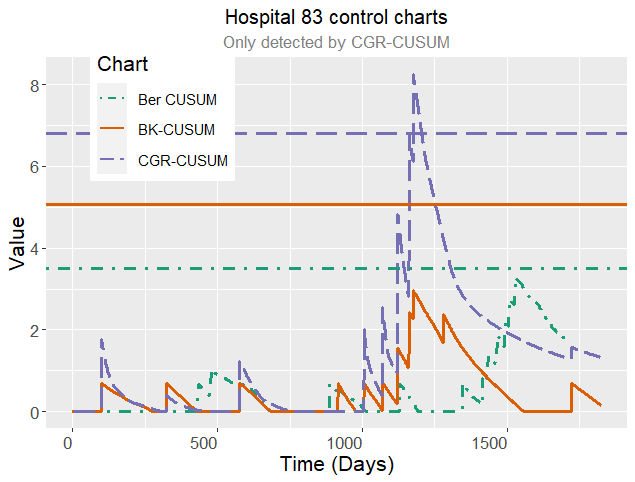}
        \caption{Hospital 83}
        \label{fig:Hosp83}
    \end{subfigure}
    \begin{subfigure}[b]{0.48\textwidth}
        \includegraphics[width=\textwidth]{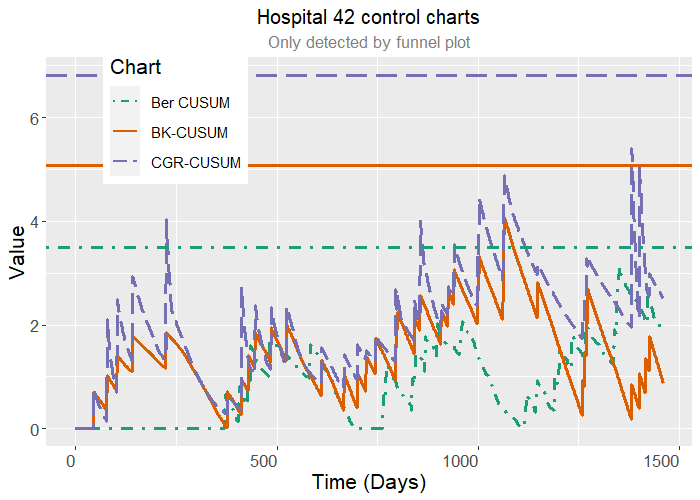}
        \caption{Hospital 42}
        \label{fig:Hosp42}
    \end{subfigure}
\caption{ The (Bernoulli) CUSUM, BK-CUSUM and CGR-CUSUM charts for four hospitals with their control limits (same color/linetype). The control limits can be found in Table \ref{table:detecttimessummary}. }
\label{fig:hospsnew}
\end{figure}

As the Bernoulli CUSUM and funnel plot both use one year post implant failure as outcome and the same risk-adjustment model, they can both detect exactly the same hospitals at the end of the third year. This is different for the continuous time CUSUM charts: both the BK- and CGR-CUSUM do not detect hospitals $5$ and $19$ and yielded one and two ``false'' detections respectively. Overall, the continuous time CUSUM procedures yield (much) faster detection times, but also signal very different hospitals than the discrete time methods, especially after the four year mark. There are multiple reasons for this. We will explain some of the possible mechanisms which cause a mismatch in detections between the methods by means of some examples in Figure \ref{fig:hospsnew}. A general observation is that the Bernoulli CUSUM has a one year delay compared to the continuous time charts. In Figure \ref{fig:Hosp5} we can see that hospital $5$ was signalled by the funnel plot and Bernoulli CUSUM but not by the BK- and CGR-CUSUM charts. Whereas the continuous time charts show a downward motion after a period of multiple consecutive failures, the Bernoulli CUSUM does not. This is most likely due to the fact that the Bernoulli CUSUM and funnel plot only consider whether an implant has failed within one year, and disregard the time of the death.  The BK-CUSUM has a similar problem, where multiple consecutive failures in a short period of time can trigger a false alarm, even if implants fail at reasonable times. A possible example of this can be seen in Figure \ref{fig:Hosp68}. We can see that the many consecutive failures make the chart jump upwards by $\log(2)$ every time, independent of the probability of failure of those implants at that point in time, thereby rapidly hitting the control limit and afterwards quickly dropping to zero. In contrast to this, the CGR-CUSUM can produce a signal when a few very unlikely failures happen in rapid succession, as can be seen in Figure \ref{fig:Hosp83}. We can see that the Bernoulli CUSUM also almost hits the control limit at a later point, as the upward jumps in the Bernoulli CUSUM chart also depend on the likelihood of failure. Finally, Figure \ref{fig:Hosp42} shows us a hospital which was only detected by the funnel plot. We can see that the hospital experiences a steady stream of failures as the value of the charts is never zero, meaning the proportion of failures at this hospital is reasonably high. The CUSUM charts however indicate that failures are happening at an acceptable rate (possibly slightly higher than target).

  The main take-away from this section is that using the continuous time methods it is possible to detect most hospitals signalled by the discrete time methods (much) faster, while guaranteeing a lower percentage of false positive signals. This follows from the fact that the type I error probability for the funnel plot was restricted to approximately $0.05$ in $3$ years, while we chose control limits for the BK- and CGR-CUSUM such that the probability of a type I error in $6$ years was $0.05$. Coincidentally, the Bernoulli CUSUM control limit of $h = 3.5$, although chosen with a different reasoning, corresponds closely to limiting the type I error in $3$ years to $0.05$ ($h = 3.3$). Additionally, the results found in this section have to be considered in the correct perspective. We cannot simply state whether the right hospitals were detected at all by any of the charts as we have no information about the true failure rates. For this reason, it is crucial to compare the performance of the charts on a set of hospitals where the true performance of the participating hospitals/implants is known. This will be done in the next section by means of comparing the (average) run length of the procedures under restrictions of the ARL and type I errors when hospitals are performing as expected.

\section{Simulation studies}
\label{sec3}

In Section \ref{sec4} we did not know which hospitals were in control.  In order to truly compare the performance of the BK- and CGR-CUSUM charts it is crucial to know which hospitals are out of control. This section will compare the methods when the true failure rates at the hospitals are known by means of simulation studies. 

In many practical applications the time to detection after problems occur is of crucial importance in monitoring the quality of the process. Therefore, comparing the performance of inspection schemes in terms of detection speed is important. The expressions found in Section \ref{sec2} for the approximate average run length of the charts provide a way to compare the charts on a theoretical basis. The equations yielded approximations, and depend strongly on the convergence rate of the maximum likelihood estimate. A simulation study can provide a better picture on the finite sample performance of said methods. Besides the detection speed, other quantities such as the type I error and power over time are of interest and will be considered later on in this section.

In Section \ref{sec4}, we chose to impose an upper limit for the MLE $e^{\hat{\theta}(t)} < 6$ in the CGR-CUSUM, due to the extremely low failure rates in the LROI data. In this section we also investigate the unrestricted CGR-CUSUM, to investigate the impact of this decision. All simulation studies performed in this section will follow the simulation procedure stated in the Supplementary materials section 6.

\subsection{A comparison of ARLs }
\label{sec:comparARL}

The main goal of this simulation section will be to compare the BK-CUSUM with the new CGR-CUSUM procedure on detection speed for out of control instances. A core assumption of the considered methods is that the change in failure rate happens instantaneously (instead of gradually) and that the true change can be quantified as a fixed increase $\exp(\theta)$ of the hazard rate. In many practical applications, both assumptions are not likely to hold. We want to examine the effect of wrongly choosing the expected change in failure rate $\exp(\theta_1)$  in the BK-CUSUM. 

To this end we consider the CGR-CUSUM and two BK-CUSUM procedures with $\exp(\theta_1) = 1.4$ and $1.8$ respectively.  We cannot use equal control limits for all charts as this would lead to different properties  under the null. For this reason we determine control limits $h$ for each procedure such that the in control ($\exp(\theta) = 1$) average run lengths of the procedures are approximately equal to $15$ years on a simulated sample size of $N = 3000$ hospitals.  For the in control hazard we use an exponential distribution with rate $\lambda = 0.002$ (time in days), so that approximately half of the subjects have failed $1$ year post procedure. We simulate patient entry by a Poisson process with rate $\psi = 2.28$ (in days), corresponding to the largest hospitals in the LROI data set. For this simulation study, risk-adjustment procedures were not considered. For the out of control situation, we want to explore what happens when the chosen $\theta_1$ is far away from the true value, so we choose true failure rates $\exp(\theta) \in \{1.2, 1.4, ..., 3\}$ to generate out of control data sets containing $N = 3000$ hospitals with the arrival rate and null hazard rate as before.

The run lengths of the two BK-CUSUM procedures are determined for each out of control data set. We determine the run length of the CGI chart on these data sets, giving us an upper bound on the run length of the CGR chart. The results can be found in Table \ref{table:simARL}, as well as the expected theoretical value of the run length as determined using equations \eqref{eq:CGIARL} and \eqref{eq:BKARL}. The calculation of the Fisher information for the exponential case is discussed in the Supplementary materials Section 7.  A notable result is that at $\exp(\theta) = 1.4$ the BK-CUSUM with $\exp(\theta_1) = 1.4$ clearly performs better than the CGR-CUSUM, but the BK-CUSUM with $\exp(\theta_1) = 1.8$  performs worse than the CGR-CUSUM. This already indicates that the impact of misspecifying  $\theta_1$ can be quite large. Surprisingly, at $\exp(\theta) = 1.8$ the CGR-CUSUM outperforms the other two charts with respect to ARL, but has the largest standard deviation in detection times. In contrast, for small values of $\theta$ the SD of the BK-CUSUM charts is larger. Finally, for very large values of $\exp(\theta) > 2$ the CGR-CUSUM seems to be the clear winner. Noticeably, the run lengths of the BK-CUSUM are way more right-skewed than those of the CGR-CUSUM. This can be explained by the non-variable ($\theta_1$) size of jumps the BK-CUSUM charts can make, in contrast to the variable ($\hat{\theta}(t)$) jump size of the CGR-CUSUM. All in all, we can conclude that with respect to detection speed the BK-CUSUM is the preferred chart when the true hazard ratio is small ($\exp(\theta) \leq 1.4$) and/or we have a lot of confidence in our prior knowledge. The approximate average run lengths determined using \eqref{eq:BKARL} and \eqref{eq:CGIARL} seem to work quite well both for the BK-CUSUM as well as for the CGR-CUSUM, especially for large ($\exp(\theta) > 1.2$) true hazard ratios. 

These simulation results give rise to the presumption that the CGR-CUSUM should perform better when the rate of failure is variable, especially combined with large values of $\theta$. This is also what we saw in Section \ref{sec4}, when we applied the CGR-CUSUM to a real-life data set.

\subsection{Power under type I error restriction}
\label{sec:powertyp1}
Instead of restricting the in control ARL, \citet{biswas} and \citet{NJR} have chosen to restrict the simulated in control type I error to $0.15$ in $5$ years and $0.1$ in $8$ years respectively. Besides this, hospitals vary in size and therefore the number of patients treated per day. This difference in patients treated per time unit, in our model expressed by the parameter $\psi$, has a strong influence on the detection speed and power of the procedures.

For this reason, in this section we will determine the power over time of two BK-CUSUM procedures ($e^{\theta_1} \in \{2, 4\}$), two CGR-CUSUM ($e^\theta \leq \{\infty, 6\}$) procedures and the Bernoulli CUSUM ($e^{\theta_1} = 2)$  for hospitals of different sizes under a restriction on the type I error. We consider $4$ groups of hospitals by size, with $\psi \in \{ 0.2, 0.6, 1, 1.7 \}$. These values were determined by subdividing the hospitals in the LROI data set  into four groups by size and averaging over their estimated patient arrival rate, see also Figure \ref{fig:Hospspsi}. Using the simulation procedure described in Supplementary materials Section 6 with resampling from the LROI data set we find control limits for all considered methods by limiting the risk-adjusted simulated type I error in $6$ years to $\alpha \approx  0.1$ on $N = 500$ in control hospitals, see Table \ref{table:hvalssimstat}. Note that the control limits for the unrestricted CGR-CUSUM are very close together for all values of $\psi$. This is a consequence of no longer bounding the MLE $\hat{\theta}(t)$ from above.



We then simulate $N = 500$ out of control ($\exp(\theta) = 2$) hospitals for each considered value of $\psi$. The detection times on these data sets are then determined for each chart using the control limits in Table \ref{table:hvalssimstat}. The resulting power over time for the BK-CUSUM ($e^{\theta_1} = 2$), the Bernoulli CUSUM and CGR-CUSUM can be seen in Figures \ref{fig:PowerCUSUM} and \ref{fig:PowerCUSUMfacet}. The BK-CUSUM with correctly specified parameters clearly has the best power over time for hospitals of all sizes. The CGR-CUSUM performs worse than the Bernoulli CUSUM for low arrival rates, but does better as the arrival rate increases. This is  due to the very high value of the control limit for the CGR-CUSUM, causing detections to be delayed.

We also compare the power over time of the BK-CUSUM ($e^\theta = 2$) with that of the CGR-CUSUM ($e^{\hat{\theta}(t)} \leq 6$) and the BK-CUSUM ($e^\theta = 4$) in Figure \ref{fig:PowerCUSUMfacet4}. In this figure, the CGR-CUSUM is clearly the winner for all hospital sizes. The control limits for the restricted CGR-CUSUM are much smaller than for the unrestricted CGR-CUSUM. This is because the unrestricted CGR-CUSUM can produce extremely large estimates for $\hat{\theta}(t)$, therefore becoming very unstable even in the in control situation. The BK-CUSUM ($e^{\theta_1} = 4$) with incorrectly specified parameters performs the worst for all hospital sizes. Notably, all $3$ procedures seem to converge towards the same power over time graph as the arrival rate increases, which was not the case in Figure \ref{fig:PowerCUSUMfacet}. We conclude that the CGR-CUSUM can yield the best power over time, but depending on the nature of the data  restricting the value of $\hat{\theta}(t)$ might be necessary to achieve such a performance.

\section{Discussion}
\label{sec5}

For almost all applications, the CGR-CUSUM will yield earlier detection times than the BK-CUSUM. This is because in any practical application the change of failure rate at a hospital will never be of a fixed size and most of the time this change will happen gradually instead of instantaneously. For this reason, the CGR-CUSUM will  perform better in any practical scenario where the expected true hazard ratio $\exp(\theta)$ is not known in advance or variable over time. This was demonstrated very clearly in our application of the charts to the LROI data set. The application of the BK- and CGR-CUSUM charts on the LROI data set also showed that in a practical application the CGR-CUSUM outperforms the BK-CUSUM with respect to detection times, while retaining a similar number of ``false'' detections. It is important to note that we do not know whether the hospitals detected by the funnel plot were the hospitals with ``true" problems, instead operating in line with \citet{SchieCUSUM} by taking the funnel plot as the golden standard.  We cannot be sure that the chosen expected hazard ratio for the BK-CUSUM was in line with reality.  All in all, we can conclude that the CGR-CUSUM is the preferred method for quality inspection, especially for large arrival rate $\psi$.

From the simulation study in Section \ref{sec:powertyp1} we concluded that the CGR-CUSUM can yield better power than the BK-CUSUM, but might require appropriately restricting the values of $\hat{\theta}(t)$. Even though the CGR-CUSUM was created with the goal of specifying less parameters, we believe that bounding $\hat{\theta}(t)$ is often more tractable than correctly specifying the expected hazard ratio. This restriction was necessary for the LROI data, but not all survival data will have an extremely low failure rate and therefore the CGR-CUSUM could also perform well without this restriction, as was seen in Section \ref{sec:comparARL}.



\subsection{Recommendations for practice}

For practical applications we suggest using the CGR-CUSUM for quality control, keeping in mind that it might be necessary to restrict the maximum likelihood estimate to an appropriate range of values (i.e. $e^{\hat{\theta}(t)} \leq 6)$. For small volume hospitals the BK-CUSUM could be preferred, as long as there is some prior information about the expected increase in failure rate. This way the small amount of information retained from patients can be partly compensated by prior knowledge. The use of a  funnel plot is not advised as it is not a real-time procedure and has the potential disadvantage of an increased risk of a type I error incurred by performing a multiple testing procedure.

We advise determining control limits $h$ for CUSUM charts either by restricting the simulated probability of a type I error over a time frame or by restricting the in control average run length of the charts. The first method may be preferred due to the lower computational requirements.

Ideally, the baseline hazard rate should be determined on a data set which is known to be in control. Realistically, this is unlikely to be feasible in many applications. The practice of considering the national average rate of failure to be in control is often sufficient. An important consideration is that any major change in the distribution of risk factors in the population will require a recalculation of control limits. Whereas information on the failure of patients can be collected in real time, the aggregation of such data over multiple hospitals is not likely to happen in real time. If the risk distribution has changed over this frame of time, it might be necessary to reconstruct the CUSUM charts, possibly leading to new or different detections.

\subsection{Limitations}

In the considered model, we assume that observations can only be right-censored. This is because in the setting of arthroplasty surgery left and interval censoring are of little interest. The same is not true for competing risks mechanisms. \citet{NJR} have considered a similar procedure to \citet{biswas} with the addition of frailty terms and competing risks, allowing for dependent competing risks. Even though they could not find an indication that the competing risks of death and revision surgery are dependent in their data, their methods can be carried over to our procedure as well. Should we be interested in detecting a decrease in the rate of failure using the BK- or CGR-CUSUM, a two-sided procedure as suggested by \citet{page} can be considered where the hypotheses of $\theta = 0$ against $\theta = \theta_1 < 0$ are used for constructing the likelihood ratio. This yields the CUSUM charts with switched positive and negative signs.


\subsection{Future work}

Additions to the CGR-CUSUM and BK-CUSUM should be considered. Notably, the power of the unrestricted CGR-CUSUM was lacking for hospitals with a low volume of patients. This is largely due to the (relatively) very high value of the control limit of the CGR-CUSUM (see Table \ref{table:hvalssimstat}). These values are so high because the CGR-CUSUM will often have an initial spike upwards when the first failure is observed due to the maximization over previous patients (i.e. all patients before the first failed patient will be ignored). When the volume of patients or failure rate is low, this leads to a large uncertainty in the determination of the MLE $\hat{\theta}(t)$. To counteract this, in Section \ref{sec4} we introduced the upper limit $\hat{\theta}(t) \leq \ln(6)$.  Another solution to the problem would be to impose a time-dependent control limit which is large at the start of the study and decreases until it reaches a fixed value, allowing the CGR-CUSUM to converge before yielding detections. A patient shuffling or weighing mechanism can be added to the CGR-CUSUM chart in order to yield quicker detection in the case of clustered failures in the past. For this, the mechanisms used by \citet{uEWMA} and \citet{STRAND} can be used as inspiration. Finally, a mechanism where patients have periods when they are not at risk of failure can be incorporated into the chart as well.

\section{Software}
\label{sec6}

Software in the form of an R package, together with a sample
input data set and complete documentation is available on CRAN at \url{https://cran.r-project.org/package=success}.

\section*{Acknowledgments}
The Dutch Arthroplasty Register (LROI) is gratefully acknowledged for providing their arthroplasty database, under agreement LROI 2020-053. This work was partially supported by NWO Veni grant $192.087$.  We wish to thank the associate editor and referees for their help in improving this manuscript.

{\it Conflict of Interest}: None declared.

\bibliographystyle{unsrtnat}
\bibliography{references}  

\begin{table}[!ht]
\resizebox{\columnwidth}{!}{\begin{tabular}{@{\extracolsep{6pt}} cccccccccc @{}}
\hline
& \multicolumn{3}{c}{\makecell{BK-CUSUM \\  $e^\theta = 1.4, h = 6.82$} } &
\multicolumn{3}{c}{\makecell{BK-CUSUM \\  $e^\theta = 1.8, h = 8.35$} }&
\multicolumn{3}{c}{\makecell{CGR/CGI \\  $ h = 7.73$}  } \\
\cline{2-4} \cline{5-7} \cline{8-10}
$e^\theta$  & ARL(SD) & MRL & Theory & ARL(SD) & MRL & Theory & ARL(SD) & MRL & Theory  \\
\hline
1 & 5510  (4930) & 4056  & $\infty$  & 5478  (4739) &  4104 &  $\infty$ &  5528  (4666) &  4398  &  $\infty$ \\
1.2& 409 (184) &374  &1352          & 639 (366) & 572 &  $\infty$         & 480 (163) &474 &  511\\
1.4& 205  (57) &198  & 227          &240 (100) &223 &  490            &229  (72) &228   &243\\
1.6& 152  (33) &148  & 159          &153  (48) &145 &  177            &153  (48) &151   &162\\
1.8& 127  (24) &125  & 130          &119  (31) &116 &  128            &117  (37) &117   &123\\
2  & 110  (20) &109  & 112          &101  (23) & 99 &  106            & 95  (30) & 94   &100\\
2.2&  99  (16) & 98  & 101          & 89  (19) & 87 &   92            & 81  (25) & 80   & 85\\
2.4&  91  (15) & 91  &  92          & 81  (16) & 80 &   82            & 71  (23) & 71   & 74\\
2.6&  85  (13) & 84  &  85          & 74  (14) & 73 &   75            & 63  (20) & 62   & 65\\
2.8&  79  (12) & 79  &  80          & 69  (13) & 68 &   70            & 57  (18) & 57   & 59\\
3  &  75  (11) & 75  &  75          & 65  (12) & 64 &   66            & 52  (17) & 51   & 54 \\
\hline
\end{tabular}}
\caption{Average/Median run length, as well as standard deviation and approximate ARL (determined using \eqref{eq:BKARL} and \eqref{eq:CGIARL}) for two BK-CUSUM with $\exp(\theta) = 1.4$ and $1.8$, as well as the CGR-CUSUM ($\exp(\theta) = 1$) and CGI-CUSUM $\exp(\theta) > 1$. Each of the quantities has been determined on a sample of $N = 3000$ hospitals with hazard ratio $\exp(\theta)$.}
\label{table:simARL}
\end{table}

\begin{table}
\begin{subtable}{\textwidth}
\centering
\begin{tabular}{@{\extracolsep{6pt}}  l c|c|c|c@{}}
\multicolumn{5}{c}{\makecell{Median (IQR) difference in detection speed (months) for  \\ hospitals detected by funnel plot in the first $3$ years} } \\
\hline
& \makecell{Funnel plot \\ $p = 0.95$ \\ yearly} & \makecell{Bernoulli CUSUM \\ $h=3.5, e^\theta = 2$ \\ monthly} & \makecell{BK-CUSUM \\ $h=5.1, e^\theta = 2$ \\ monthly} & \makecell{CGR-CUSUM \\ $h=6.8, e^{\hat{\theta}(t)} \leq 6$ \\ monthly} \\
\hline
Funnel plot     &   0 (0 - 0)    &  9 (6 - 12)     & 15 (12 - 17.5)        &17 (15 - 18) \\
Bernoulli CUSUM      &   &     0 (0 - 0)   &     7 (4.5 - 8)       &9 (5 - 10) \\
BK-CUSUM    & &      &     0 (0 - 0)     &    1 (-0.5 - 3) \\
CGR-CUSUM   & &      &    &         0 (0 - 0) \\
\end{tabular}
\end{subtable}
\caption{Difference in detection speed (months) of columns with respect to rows. Positive indicating quicker detection and negative indicating slower detection speeds. Values determined on hospitals detected by the funnel plot in the first $3$ years, with missing detections omitted.  }
\label{table:detecttimessummary}
\end{table}

\begin{table}[!ht]
\centering
\begin{tabular}{@{\extracolsep{6pt}} cccccc  @{}}
 & \multicolumn{4}{c}{Control limit h} \\
\cline{2-6}
 & Bernoulli CUSUM  & BK-CUSUM & BK-CUSUM & CGR-CUSUM & CGR-CUSUM \\
$\psi$ & $e^{\theta_1} = 2$ & $e^{\theta_1} = 2$ & $e^{\theta_1} = 4$ & N/A & $e^{\hat{\theta}(t)} \leq 6$ \\
\cline{1-1} \cline{2-2} \cline{3-3} \cline{4-4} \cline{5-5} \cline{6-6}
0.2 & 2.62 & 3.15 & 4.64 & 7.31 & 4.68 \\ 
  0.6 & 3.71 & 4.19 & 5.81 & 7.73 & 5.79 \\ 
  1 & 4.34 & 4.76 & 6.34 & 8.27 & 6.51 \\ 
  1.7 & 4.72 & 5.41 & 6.79 & 8.54 & 6.69 \\ 
\end{tabular}
\caption{Control limits determined on a sample size of $N = 500$ in control ($e^\theta = 1$) hospitals such that the type I error in $6$ years  $\alpha \approx 0.1$. }
\label{table:hvalssimstat}
\end{table}

\begin{figure}[!ht]
\centering
 \begin{subfigure}[b]{0.48\textwidth}
        \centering
        \includegraphics[width=\textwidth]{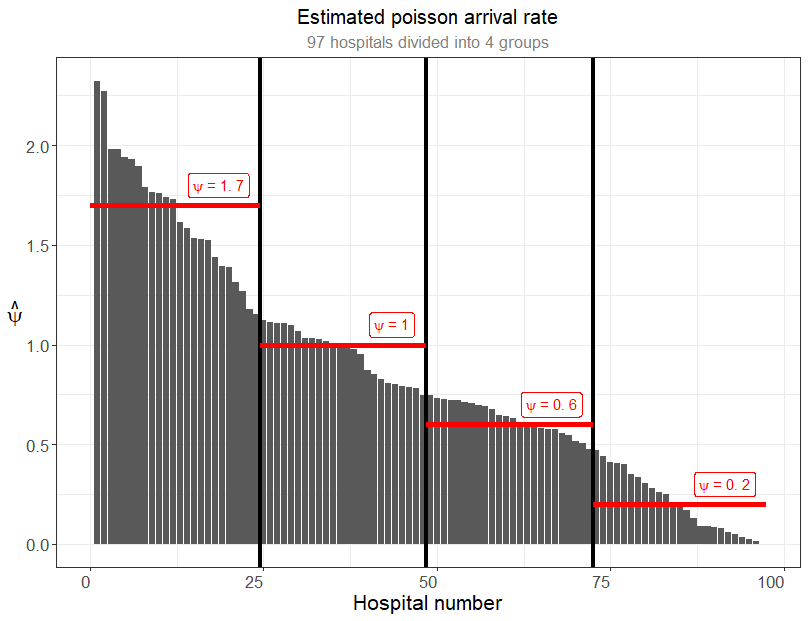}
        \caption{Estimated arrival rate}
        \label{fig:Hospspsi}
    \end{subfigure}
    \begin{subfigure}[b]{0.48\textwidth}
    \centering
        \includegraphics[width=\textwidth]{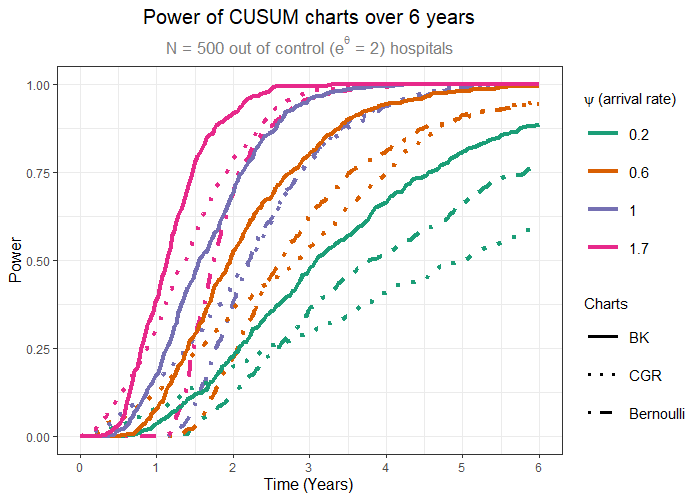}
        \caption{Power over time}
        \label{fig:PowerCUSUM}
    \end{subfigure}
        \begin{subfigure}[b]{0.48\textwidth}
        \includegraphics[width=\textwidth]{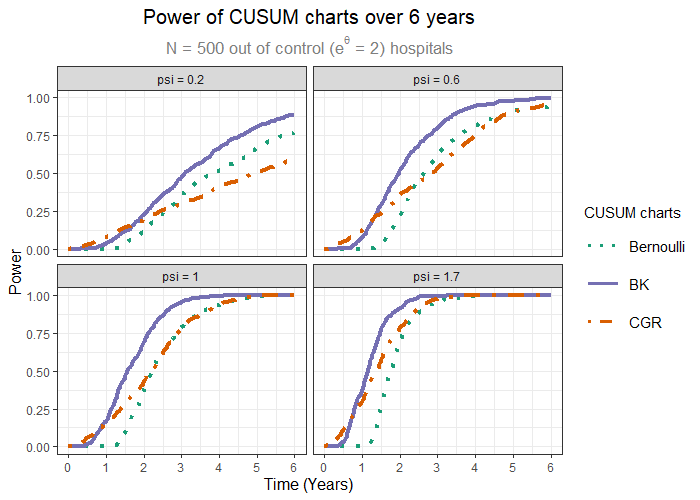}
        \caption{Power over time}
        \label{fig:PowerCUSUMfacet}
    \end{subfigure}
        \begin{subfigure}[b]{0.48\textwidth}
        \includegraphics[width=\textwidth]{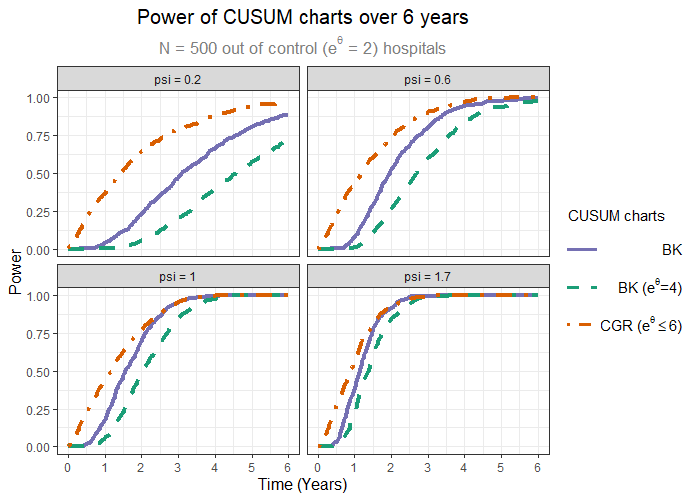}
        \caption{Power over time}
        \label{fig:PowerCUSUMfacet4}
    \end{subfigure}
\caption{ (a) Estimated arrival rate as well as the subdivision of the hospitals into $4$ groups. (b) Simulated power of the Bernoulli and continuous time CUSUM charts on a sample size of $N = 500$ out of control ($\exp(\theta_1) = 2)$ hospitals using control limit values such that the simulated in control type I error  $\alpha \approx 0.1$ in $6$ years. (see Table \ref{table:hvalssimstat}) (c) Figure (b) faceted over the different values of $\psi$. (d) Comparison of the power over time of two BK-CUSUM charts ($\exp(\theta_1) \in \{ 2, 4 \})$ and the CGR-CUSUM with $\exp(\hat{\theta}(t)) \leq 6$.  }
\label{fig:powerenzo}
\end{figure}






\end{document}


\maketitle

\section{Some proofs}
To prove our main result stated in Theorem \ref{theorem:mainresult}, we first need to derive some properties of other quantities, which will be done in this section.

\begin{lem}
\label{lem:EA(t)}
Assume that $f_i^\theta$ and $h_i^\theta$ are non-negative and Borel measurable. Define $\Lambda(t) = \sum_{i \geq 1} \Lambda_i(t)$. Then:
\begin{align*}
\mathbb{E}[d\Lambda(u)] &= e^{-\theta}  \psi \mathbb{E}_{Z_i} \left[  F_i^\theta(u)  \right] du \\ 
&=: \gamma_u du
\end{align*} and:
\begin{align*}
\mathbb{E}[\Lambda(t)] &= \int_0^t \gamma_u du
\end{align*}

\newpage
with $\psi$ the rate of arrivals and $e^\theta \Lambda_i(t)$ the true risk-adjusted subject specific cumulative intensity of failure at the institute of interest.
\end{lem} 
\begin{proof}
We consider a hospital with hazard rate $e^\theta$ times the baseline hazard rate. We define:
\begin{align}
\label{UGLRTINAPPROX}
U_{CGI}(t) &:= \hat{\theta}(t) N(t) + (e^{\hat{\theta}(t)} -1) \Lambda(t)
\end{align} where: 
\begin{align*}
\hat{\theta}(t) &= \max \left\lbrace 0, \ln \left(  \frac{N(t)}{ \Lambda(t)} \right) \right\rbrace.
\end{align*}

Assume that patients arrive according to a homogeneous Poisson process with rate $\psi > 0$. We choose to consider the lifetime of all patients up until the time of failure (or censoring). The first step is to calculate the  expected value of $d\Lambda(u)$:
\begin{align*}
\mathbb{E}[d\Lambda(u)] &= \mathbb{E}\left[  \sum_{i \geq 1} Y_i(u) h_i(u-S_i) du  \right] \\
&= e^{-\theta} \mathbb{E} \left[  \sum_{i \geq 1} \mathbbm{1}\{ S_i \leq u , X_i \geq u-S_i  \} e^\theta h_i(u-S_i) du  \right].
\end{align*}
Here we use that $T_i = S_i + X_i$. Then using the law of total expectation twice (conditioning first on $Z_i$ and then on $S_i$):
\begin{align*}
\mathbb{E}[d\Lambda(u)] &= e^{-\theta} \sum_{i \geq 1} \mathbb{E}_{S_i} \left[ \mathbb{E}_{Z_i} \left[ \mathbb{E} \left[  \mathbbm{1}\{ S_i \leq u \} \mathbbm{1} \{ X_i \geq u - S_i  | Z_i \} e^\theta h_i(u-S_i | Z_i) du  \right] | S_i \right] \right] \\
&= e^{-\theta} \sum_{i \geq 1} \mathbb{E}_{S_i} \left[ \mathbbm{1}\{ S_i \leq u \}  \mathbb{E}_{Z_i} \left[ \mathbb{E} \left[   \mathbbm{1} \{ X_i \geq u - S_i | Z_i \}  \right] e^\theta h_i(u-S_i | Z_i) du  | S_i \right] \right] \\
&= e^{-\theta} \sum_{i \geq 1} \mathbb{E}_{S_i} \left[ \mathbbm{1}\{ S_i \leq u\}  \mathbb{E}_{Z_i} \left[    \mathbb{P} \left( X_i \geq u - S_i | Z_i \right)   h_i^\theta(u - S_i | Z_i) du  | S_i \right] \right] \\
&= e^{-\theta} \sum_{i \geq 1} \mathbb{E}_{S_i} \left[ \mathbbm{1}\{ S_i \leq u \}  \mathbb{E}_{Z_i} \left[    S_i^\theta \left(  u-S_i  | Z_i \right)   h_i^\theta(u-S_i | Z_i) du  | S_i \right] \right] \\
&= e^{-\theta} \sum_{i \geq 1} \mathbb{E}_{S_i} \left[ \mathbbm{1}\{ S_i \leq u\}  \mathbb{E}_{Z_i} \left[    f_i^\theta (u-S_i | Z_i)   | S_i \right] \right]du  \\
&= e^{-\theta} \sum_{i \geq 1} \int_0^{u} \mathbb{E}_{Z_i} \left[    f_i^\theta (u - x )    \right] \psi \frac{e^{-\psi x}(\psi x)^{i-1}}{(i-1)!} dx du \\
&= e^{-\theta}  \psi \int_0^{u} \mathbb{E}_{Z_i} \left[    f_i^\theta (u - x )    \right] e^{-\psi x} \sum_{i \geq 1} \frac{(\psi x)^{i-1}}{(i-1)!} dx du. \\
\end{align*}  We obtain using Fubini's Lemma that:
\begin{align}
\label{dAuGLR}
\mathbb{E}[d\Lambda(u)] &= e^{-\theta}  \psi \int_0^{u} \mathbb{E}_{Z_i} \left[    f_i^\theta (u - x )    \right]  du \\
&= e^{-\theta}  \psi \mathbb{E}_{Z_i} \left[  F_i^\theta(u)  \right] du.
\end{align} We define, similarly to above, the notation:
\begin{align}
\label{gammamu}
\gamma_u &:= e^{-\theta} \psi \mathbb{E}_{Z_i}[F_i^\theta(u)] > 0.
\end{align} We assumed that $f_i^\theta$ and $h_i^\theta$ are non-negative and Borel measurable, therefore using Fubini's Lemma once again we obtain that $\mathbb{E}[\Lambda(t)] = \int_0^t \gamma_u du$.
\end{proof}

\section{Fisher Information}
\label{sec:fisherinfo}

We derive the Fisher information contained in all observations:

\begin{lem}
\label{lem:fishertheta}
The Fisher information in all observations at time $t >0$ is given by:
\begin{align}
I(\theta, t) &= \psi \int_0^t \mathbb{E}_{Z_i} \left[ F_i^\theta(k) \right] dk.
\label{fischerinfotheta}
\end{align}
\end{lem}
\begin{proof}
The likelihood function for all observations is given by:
\begin{align*}
L(\theta|t) &=  \prod_{i \geq 1} \left( e^\theta d\Lambda_i(t) \right)^{dN_i(t) } e^{-e^\theta \Lambda_i(t)}.
\end{align*}  The log likelihood ratio is then given by:
\begin{align*}
l(\theta|t) &=  \sum_{i \geq 1}(dN_i(t) )(\theta + \ln(d\Lambda_i(t)))  - e^\theta  \sum_{i \geq 1} \Lambda_i(t).
\end{align*} Taking the derivative w.r.t. $\theta$ yields:
\begin{align*}
\frac{\partial l(\theta|t)}{\partial \theta} &= \sum_{i \geq 1} dN_i(t)  - e^\theta  \sum_{i \geq 1}\Lambda_i(t).
\end{align*} And the second derivative:
\begin{align*}
\frac{\partial^2 l(\theta|t)}{\partial \theta^2} &= - e^\theta \sum_{i \geq 1} \Lambda_i(t).
\end{align*} Finally, the Fisher information is given by:
\begin{align}
I(\theta, t) &=  - \mathbb{E} \left[ - e^\theta \sum_{i \geq 1}  \Lambda_i(t) \left. \right| \theta  \right]  \\
&= e^\theta  \mathbb{E}_\theta \left[  \Lambda(t) \right] \\
&= e^\theta  e^{-\theta } \psi \int_0^t \mathbb{E}_{Z_i} \left[ F_i^\theta(k) \right] dk \\
&= \psi \int_0^t \mathbb{E}_{Z_i} \left[ F_i^\theta(k) \right] dk
\end{align} where we have used our result from Lemma \ref{lem:EA(t)}.
\end{proof}

\section{Convergence of MLE}

In this section we reiterate a result from  \citet{hoadley} and transform it so it becomes usable in our problem at hand.

\begin{lem}
\label{lem:convthetat}
Let $t > 0$ and suppose we have $n$ patients. Let $\{f_i^\theta, i = 1, ..., n\}$ meet conditions (N1)-(N9) of \cite{hoadley}. Then, as $n \to \infty$:
\begin{align*}
\sqrt{n}(\hat{\theta}(t) - \theta) \overset{d}{\to} \mathcal{N} \left( 0, \frac{1}{\overline{I(\theta, t)}}\right)
\end{align*} with $\overline{I(\theta, t)} = \frac{I(\theta, t)}{n}$ and $I(\theta, t)$ the Fisher information in all observations at time $t$. Moreover, assuming that $n = \psi \cdot t$ we have that:
\begin{align*}
\sqrt{ t}\left( \hat{\theta}(t) - \theta \right) \overset{d}{\to} \mathcal{N} \left(0, \frac{1}{\psi \cdot \overline{I(\theta, t)}} \right)
\end{align*} as $t \to \infty$.
\end{lem}
\begin{proof}
Section $4$ of  \citep{hoadley} tells us that under conditions $(N1)-(N9)$ as stated in the article our first statement holds. Most of these conditions are likely to hold when $h_i^\theta$ is continuous and twice differentiable. Finally, as patients arrive according to a Poisson process with rate $\psi$, it is reasonable (especially when $\psi$ is large) to assume that $n = t \cdot \psi$, as we expect to see $\psi$ patients arrive per time unit. Using this relation we then have that $n \to \infty$ implies that $t \to \infty$ (remember that $\psi$ is constant), therefore using the properties of a normal distribution we obtain the second statement.
\end{proof}

\section{Proof of Theorem 2.1}
In this section we prove the main theorem in our article. This proof uses the results stated in the previous sections.

\begin{theorem}
\label{theorem:mainresult}
Suppose that patients arrive according to a Poisson process with rate $\psi$ and let some other regularity conditions hold. It follows that when $\theta > 0$:
\begin{align*}
    \sqrt{t} \left( CGI(t) - (\theta + \exp(-\theta) -1) I(\theta, t) \right) \overset{d}\rightarrow \mathcal{N} \left( 0, t \theta^2 I(\theta, t)  \right)
\end{align*} and when $\theta = 0$:
\begin{align*}
    t \cdot CGI(t) \overset{d}\rightarrow \Gamma \left(0.5, t\right)
\end{align*} where 
\begin{align*}
    I(\theta, t) = \psi \int_0^t \mathbbm{E}_{Z_i} \left[  F_i^\theta(k) \right] dk 
\end{align*}is the Fisher information in all observations at time $t$.
\end{theorem}
\begin{proof}
\textbf{1. When }\bm{$\theta > 0$}\textbf{:}  Assuming that $N(t) = e^{\hat{\theta}(t)} \Lambda(t)$ we can write:
\begin{align*}
CGI(t) &= \hat{\theta}(t) N(t) - (e^{\hat{\theta}(t)} -1) \Lambda(t) \\
&= \hat{\theta}(t) e^{\hat{\theta}(t)} \Lambda(t) - (e^{\hat{\theta}(t)} -1) \Lambda(t).
\end{align*} Now assume that $\Lambda(t)$ is a constant with value $\mathbb{E}[\Lambda(t)] =\int_0^t \gamma_u du = e^{-\theta} I(\theta, t)$ (using Lemma \ref{lem:EA(t)}). We can write:
\begin{align*}
CGI(t) &= \hat{\theta}(t) e^{\hat{\theta}(t)} e^{-\theta} I(\theta, t) - (e^{\hat{\theta}(t)} -1) e^{-\theta} I(\theta, t).
\end{align*} Then consider the function $\phi$ defined as:
\begin{align*}
\phi(x) &= x e^{x} e^{-\theta} I(\theta, t) - e^x e^{-\theta} I(\theta, t) + e^{-\theta} I(\theta, t)
\end{align*} and note that $\phi(\hat{\theta}(t)) = CGI(t)$. Additionally, we find that $\phi$ is differentiable at $\theta$ with derivative:
\begin{align*}
\phi'(x) &= e^{-\theta} I(\theta, t) xe^x.
\end{align*} Then by the delta method (see section $7$ of  \citet{asympvaart}) and the result of Lemma \ref{lem:convthetat} we obtain that:
\begin{align*}
\sqrt{n} \left( \phi(\hat{\theta}(t)) - \phi(\theta)  \right) \overset{d}{\rightarrow} \mathcal{N}\left(0, \frac{\left(\phi'(\theta)\right)^2}{\overline{I(\theta, t)}}  \right)
\end{align*} as $n \to \infty$ which reduces to:
\begin{align*}
\sqrt{t} \left( CGI(t) - \left(\theta + e^{-\theta} -1 \right) I(\theta, t) \right) \overset{d}{\rightarrow} \mathcal{N}\left( 0,t  \theta^2 I(\theta, t) \right)
\end{align*} when $t \to \infty$ using a similar argument ($n = \psi \cdot t$) as in the proof of Lemma \ref{lem:convthetat}.

\textbf{2. When }\bm{$\theta = 0$}\textbf{:} In this case we can no longer use the delta method to determine the distribution of $CGI(t)$ as $\phi'(\theta) = \phi'(0) = 0$. Luckily we can use the second-order delta method (see Theorem $5.5.26$ of \citet{delta2nd}). Note that:
\begin{align*}
\phi''(x) &= (x+1) e^x e^{-\theta}I(\theta, t)
\end{align*} and $\phi''(0)  = I(\theta, t)$. Now the second-order delta-method in combination with Lemma \ref{lem:convthetat} tells us that:
\begin{align*}
n \left(  CGI(t) - \phi(0)\right) \overset{d}{\rightarrow} \frac{1}{\overline{I(\theta,t)}} \frac{\phi''(0)}{2} \chi^2_1
\end{align*} as $n \to \infty$ which simplifies to:
\begin{align*}
t \cdot CGI(t) \overset{d}{\rightarrow} \frac{t}{2} \chi^2_1 = \Gamma \left( \eta = \frac{1}{2}, \nu = t \right)
\end{align*} as $t \to \infty$, using the shape/scale($\eta/\nu$) parametrization of the Gamma distribution (and using that $n = \psi \cdot t$).
\end{proof}

\subsection{A martingale approach}
A result similar to Theorem \ref{theorem:mainresult} can be proved using a Martingale approach.

\begin{theorem}
As $t$ becomes large, the expected value of the CGI-CUSUM chart is approximated by:
\begin{align*}
    \mathbb{E}[CGI(t)] \approx (\theta e^\theta - e^\theta + 1) \Lambda(t) 
\end{align*}
\end{theorem}
\begin{proof}

The CGR-CUSUM can be written as:
\begin{align}
CGI(t) &= \hat{\theta}(t) N(t) - \left( e^{\hat{\theta}(t)} -1\right) \Lambda(t) \\
&= \left( \hat{\theta}(t) - \theta \right) N(t)  - \left( e^{\hat{\theta}(t)} -e^\theta \right) \Lambda(t) + \theta N(t) - e^\theta \Lambda(t) + \Lambda(t)
\label{eq:martCGR}
\end{align}

Lemma \ref{lem:convthetat} tells us that $\sqrt{t} \left( \hat{\theta}(t) - \theta \right)$ converges to a normal distribution with mean zero and strictly decreasing variance $\frac{1}{\psi \overline{I(\theta, t)}}$ as $t$ becomes large. Applying the delta method we obtain that $\sqrt{t} \left( e^{\hat{\theta}(t)} -e^\theta \right)$ converges to a normal distribution with mean zero and strictly decreasing variance $\frac{e^{2\theta}}{\psi \overline{I(\theta, t)}}$ as $t$ becomes large. As $t$ becomes large, the first two terms of Equation \eqref{eq:martCGR} will therefore become small. We can then approximate:
\begin{align*}
CGI(t) \approx \theta N(t) - e^\theta \Lambda(t) + \Lambda(t)
\end{align*} When $\theta > 0$, $e^\theta \Lambda(t)$ is the compensator of $N(t)$  and $M(t) := N(t) - e^\theta \Lambda(t)$ is a zero-mean martingale as $M(0) = 0$ per construction (see \cite{Aalen_2011} Section 2.2.5). We rewrite CGR$(t)$:
\begin{align*}
CGI(t) &\approx \theta N(t) -  \theta e^\theta \Lambda(t) + \theta e^\theta \Lambda(t) - e^\theta \Lambda(t) + \Lambda(t) \\
&= \theta M(t) + (\theta e^\theta - e^\theta + 1) \Lambda(t) 
\end{align*} Again, we assume $\Lambda(t)$ is constant with value $\mathbb{E}[\Lambda(t)]  = e^{-\theta} I(\theta, t)$ (using Lemma \ref{lem:EA(t)}) to obtain:
\begin{align*}
CGI(t) &\approx \theta M(t) + (\theta e^\theta - e^\theta + 1) e^{-\theta} I(\theta, t) \\
&= \theta M(t) + (\theta + e^{-\theta} - 1 ) I(\theta, t)
\end{align*} Then conditioning on the history at time zero $\mathcal{F}_0$ we obtain:
\begin{align*}
\mathbb{E} \left[ CGI(t)  | \mathcal{F}_0 \right] &= \theta \cdot 0 +  (\theta + e^{-\theta} - 1 ) I(\theta, t) \\
&= (\theta + e^{-\theta} - 1 ) I(\theta, t)
\end{align*} where we have used the martingale property $\mathbb{E}[M(t) | \mathcal{F}_s] = M(s)$ for all $s \leq t$.
\end{proof}

Equating this expression to $h > 0$ and solving for $t$ we obtain the same expression for the average run length of the CGR-CUSUM as using the method above.

\section{Approximate ARL of BK-CUSUM}

\begin{corollary}
Suppose $\theta_1$ is chosen such that $\exp(\theta_1)/\exp(\theta) < \theta_1 + \exp(-\theta)$. We find an approximate average run length $\widehat{\mathrm{ARL}}_{BK}(\theta, h)$ by solving the following equation for $t$:
\begin{align}
    \left(\theta_1 + \exp(-\theta) - \frac{\exp(\theta_1)}{\exp(\theta)}\right) I(\theta, t) = h.
    \label{eq:BKARL}
\end{align}
\end{corollary}
\begin{proof}
Assuming that $N(t) = e^{\theta} \Lambda(t)$ we can write:
\begin{align*}
BK(t) &= \theta_1 N(t) - (e^{\theta_1} -1) \Lambda(t)\\
&= \theta_1 e^\theta \Lambda(t) - (e^{\theta_1} -1) \Lambda(t).
\end{align*} From Lemma \ref{lem:EA(t)} we know that $\mathbb{E}[\Lambda(t)] =\int_0^t \gamma_u du = e^{-\theta} I(\theta, t)$. Taking the expected value of $BK(t)$ we then obtain:
\begin{align*}
\mathbb{E}\left[ BK(t) \right] &=  \theta_1 e^\theta e^{-\theta} I(\theta, t) - (e^{\theta_1} -1) e^{-\theta} I(\theta, t)\\
&= ( \theta_1 + e^{-\theta} - \frac{e^{\theta_1}}{e^\theta}) I(\theta, t)
\end{align*}

By equating the expected value of this expression to $h$ we can find an approximate average run length for the BK-CUSUM. As $I(\theta, t)$ is non-negative, this is only possible for control limits $h > 0$ when $( \theta_1 + e^{-\theta} - \frac{e^\theta_1}{e^\theta}) > 0$.
\end{proof}

\section{Standard simulation procedure}
\label{simprocedure} \hfill
\begin{itemize}
\item Step 1: Generating a training (in control) data set with $N$ hospitals.
\begin{enumerate}
\item Choose (parametric) null cumulative baseline hazard rate or determine from existing data (for example, using R package survival \citep{survivalR}).
\item Generate patient arrival times in the required time frame using Poisson arrivals with rate $\psi$.
\item (Optional) Resample patient characteristics from data set.
\item Determine (risk-adjusted) survival times for every patient using above chosen cumulative baseline hazard rate using the method described by \citep{Gensurv}.
\item Repeat 2-4 N times. Combine into single data set.
\end{enumerate}
\item Step 2: Determining a suitable control limit $h$.
\begin{enumerate}
\item Determine a (parametric) cumulative baseline hazard rate using the generated in control data set. Optionally, use the cumulative hazard rate from step $1$.
\item Construct the charts on the training data set.
\item Determine control limit $h$ such that required restrictions (on sensitivity or ARL under the null) are met for the collection of the constructed charts.
\end{enumerate}
\item Step 3: Generate test (out of control) data set.
\begin{enumerate}
\item Follow step $1$ with $\theta > 0$ as required.
\end{enumerate}
\item Step 4: Evaluate charts on test data set
\begin{enumerate}
\item Construct the charts on the test data with the control limit determined in step $2$. 
\item Approximate required quantities (such as ARL, sensivity, specificity) from these charts.
\end{enumerate}
\end{itemize}

\section{Calculation of the Fisher information}

There are two hurdles in the calculation of the Fisher information found in Section \ref{sec:fisherinfo}. First of all we need to calculate $\mathbb{E}_{\mathbf{Z}_i} \left[ F_i^\theta(s) \right]$. We propose a method to tackle this calculation in Section \ref{sec:RAfisher} as well as some examples for the PVF family of distributions. The second hurdle is then calculating the resulting integral $\int_0^t \mathbb{E}_{\mathbf{Z}_i} \left[ F_i^\theta(s) \right] ds$. For this step, an assumption must be made for the hazard rate. As an example, we calculate a closed form expression for the Fisher information for exponential failure times in Section \ref{sec:expFisher}.

\subsection{Risk-adjustment}
\label{sec:RAfisher}

An approach which can be used to calculate the expected value of the risk-adjusted cumulative distribution function is by using Laplace transforms.

\begin{lem}
\label{lem:laplacefisher}
Assuming that $e^{\mathbf{Z}_i^\top \pmb{\beta}} \sim U$ with $U$ a general distribution, the Fisher information at time $t$ is given by:
\begin{align*}
I(\theta, t) &= \psi t - \psi \int_0^t \mathcal{L}(H^\theta(s)) ds
\end{align*} with $\mathcal{L}(H^\theta(s)) = \mathbb{E} \left[  e^{-H^\theta(s)U} \right]$ the Laplace transform of $U$ and $H^\theta(k) = e^\theta H(k)$ the cumulative hazard rate multiplied by the hazard ratio $e^\theta$.
\end{lem}
\begin{proof}
Note that we can write:
\begin{align*}
I(\theta, t) &= \psi \int_0^t \mathbb{E}_{\mathbf{Z}_i} \left[ F_i^\theta(s) \right] ds \\
&= \psi  \int_0^t \mathbb{E}_{\mathbf{Z}_i} \left[ 1- S_i^\theta(s) \right] ds \\
&= \psi t - \psi  \int_0^t \mathbb{E}_{\mathbf{Z}_i} \left[ e^{-H^\theta(s) e^{\mathbf{Z}_i^\top \pmb{\beta}}} \right] ds \\
\end{align*} with $H^\theta(k) = e^\theta H(k)$ the cumulative hazard rate multiplied by the hazard ratio $e^\theta$. We assumed that  $e^{\mathbf{Z}_i^\top \pmb{\beta}} \sim U$ where $U$ follows some general distribution. Then define the Laplace transform of $U$ as follows:
\begin{align*}
\mathcal{L}(c) = \mathbb{E} \left[  e^{-cU} \right]
\end{align*} We can then write:
\begin{align*}
I(\theta, t) &= \psi t - \psi \int_0^t \mathcal{L}(H^\theta(s)) ds
\end{align*}
\end{proof}

\begin{Example}[Example]{Gamma distribution \\}
A common assumption is to take $U \sim \Gamma(\eta, \nu)$ (shape/scale parametrization). We then obtain:
\begin{align*}
I(\theta, t) &= \psi t - \psi \int_0^t \left( \frac{\nu}{\nu + H^\theta(s)} \right)^\eta ds
\end{align*} Taking the mean of the covariate distribution to be equal to $1$ (i.e. $\eta = \nu$) and defining $\delta = \frac{1}{\nu}$ we then obtain:
\begin{align*}
I(\theta, t) &= \psi t - \psi \int_0^t \left( 1 + \delta H^\theta(s) \right)^{-\frac{1}{\delta}} ds
\end{align*}
\end{Example}

\begin{Example}[Example]{Family of PVF distributions \\}
Consider the family of PVF distributions, which are distributions having a Laplace transform in the form:
\begin{align*}
\mathcal{L}(c; \rho, \nu, m) &= \exp \left[-\rho \left\{ 1 - \left( \frac{\nu}{\nu + c} \right)^m  \right\}  \right]
\end{align*} with $\nu > 0$, $m > -1$ and $m \rho > 0$. For more information on this family of distributions, we direct the reader to \cite{Aalen_2011} Section 6.2.3. We then obtain the following expression for the Fisher information:
\begin{align*}
I(\theta, t) &= \psi t - \psi \int_0^t \exp \left[-\rho \left\{ 1 - \left( \frac{\nu}{\nu + H^\theta(s)} \right)^m  \right\}  \right] ds
\end{align*} 
\end{Example}

\subsection{Exponential failure times}
\label{sec:expFisher}
In this section we consider the case when the failure times are exponentially distributed.

\begin{Example}[Example]{Exponential distribution (no covariates)\\}
Suppose failure times are exponentially distributed with parameter $\lambda > 0$. We calculate the Fisher information $I(\theta, t)$ without risk-adjustment:
\begin{align*}
    I(\theta, t) &= \psi \int_0^t F^\theta(s) ds \\
    &= \psi \int_0^t 1 - e^{-H^\theta(s)} ds = \psi \int_0^t 1 - e^{-H(s) e^\theta} ds \\
    &= \psi \int_0^t 1 - e^{-\lambda s e^\theta} ds \\
    &= \psi t - \psi \left[ \frac{e^{-\lambda e^\theta s}}{\lambda e^\theta}  \right]^{t}_0 \\
    &= \psi \left( t - \frac{1-e^{-\lambda e^\theta t}}{\lambda e^\theta} \right)
\end{align*} with $H(s)$ the cumulative hazard rate.
\end{Example}

\begin{Example}[Example]{Exponential distribution with Gamma distributed risk-adjustment\\}
Suppose failure times are exponentially distributed with parameter $\lambda > 0$. Using Lemma \ref{lem:laplacefisher} and assuming that the covariates are Gamma distributed with mean $1$ and variance $\delta$ we find:
\begin{align*}
I(\theta, t) &= \psi t - \psi \int_0^t \left( 1 + \delta \lambda s e^\theta \right)^{-\frac{1}{\delta}} ds \\
&= \psi t - \psi \left[ \frac{\left( 1 + \delta \lambda e^\theta s \right)^{1-\frac{1}{\delta}}}{\lambda e^\theta \left( \delta -1 \right)}  \right]_0^t \\
&= \psi t - \psi \left( \frac{\left( 1 + \delta \lambda e^\theta t \right)^{1-\frac{1}{\delta}} - 1}{\lambda e^\theta (\delta -1)}  \right)
\end{align*}
\end{Example}

\subsection{Numerical computation}

For covariate distributions which do not have an (easy) Laplace transform or hazard rates with no closed form expression, we suggest using the relationship:
\begin{align*}
    F^{\theta}_{i}(t) = 1 - S_{i}^{\theta}(t) = 1 - e^{-H(t) \exp{(\theta)} \exp{\left( \mathbf{Z}_i^\top \pmb{\beta} \right)}}
\end{align*} with $H(t)$ the cumulative hazard rate. The resulting function can then be numerically integrated first with respect to $\exp{\left( \mathbf{Z}_i^\top \pmb{\beta} \right)}$ and then with respect to $t$ to obtain $I(\theta, t)$. Without risk-adjustment, $\exp{\left( \mathbf{Z}_i^\top \pmb{\beta} \right)}$ can be left out of the equation.




\section{Additional Figures and Tables}

\begin{figure}
    \centering
    \begin{subfigure}[b]{0.48\textwidth}
    \includegraphics[width = \textwidth]{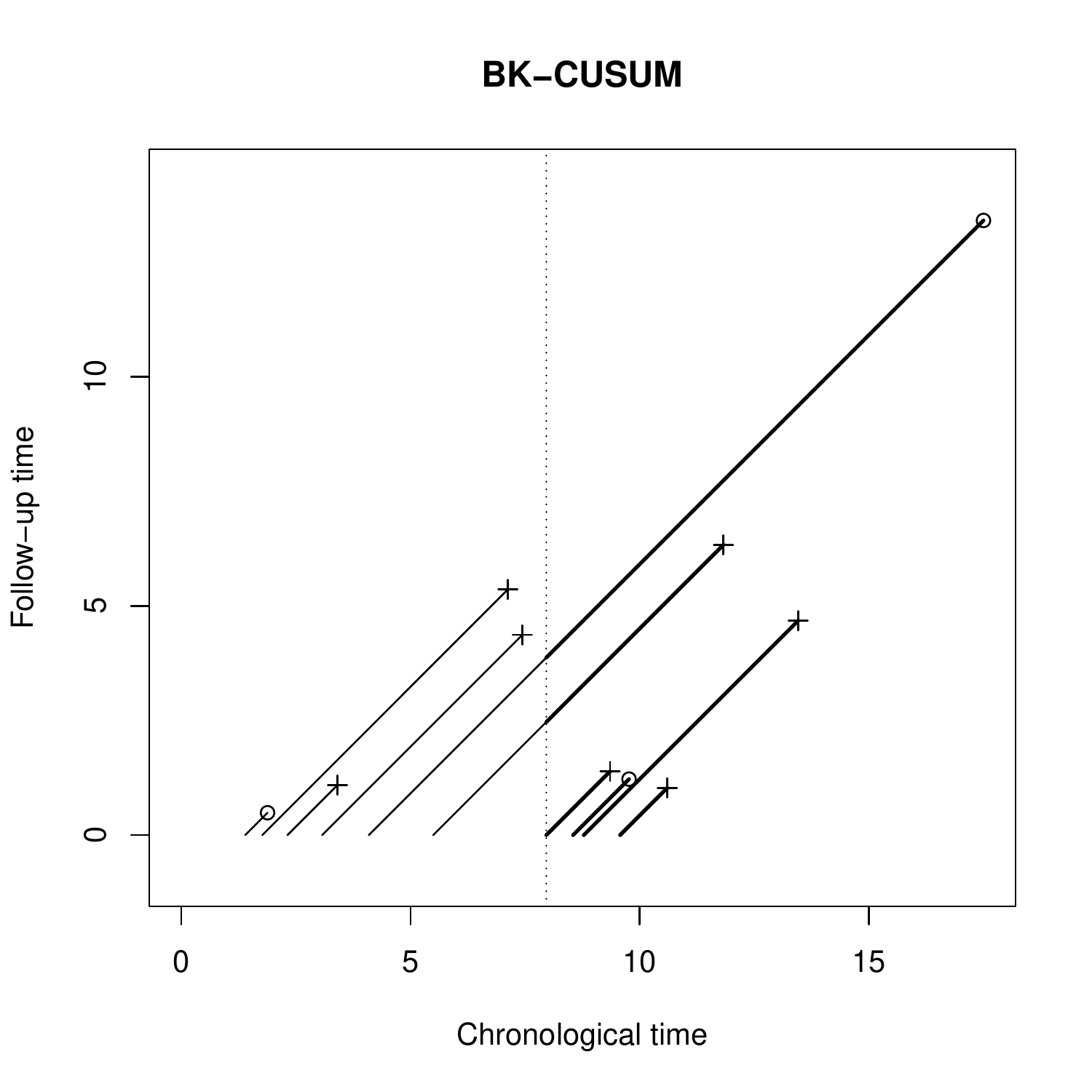}
    \caption{Lexis diagram of BK-CUSUM}
    \end{subfigure}
    \begin{subfigure}[b]{0.48\textwidth}
    \includegraphics[width = \textwidth]{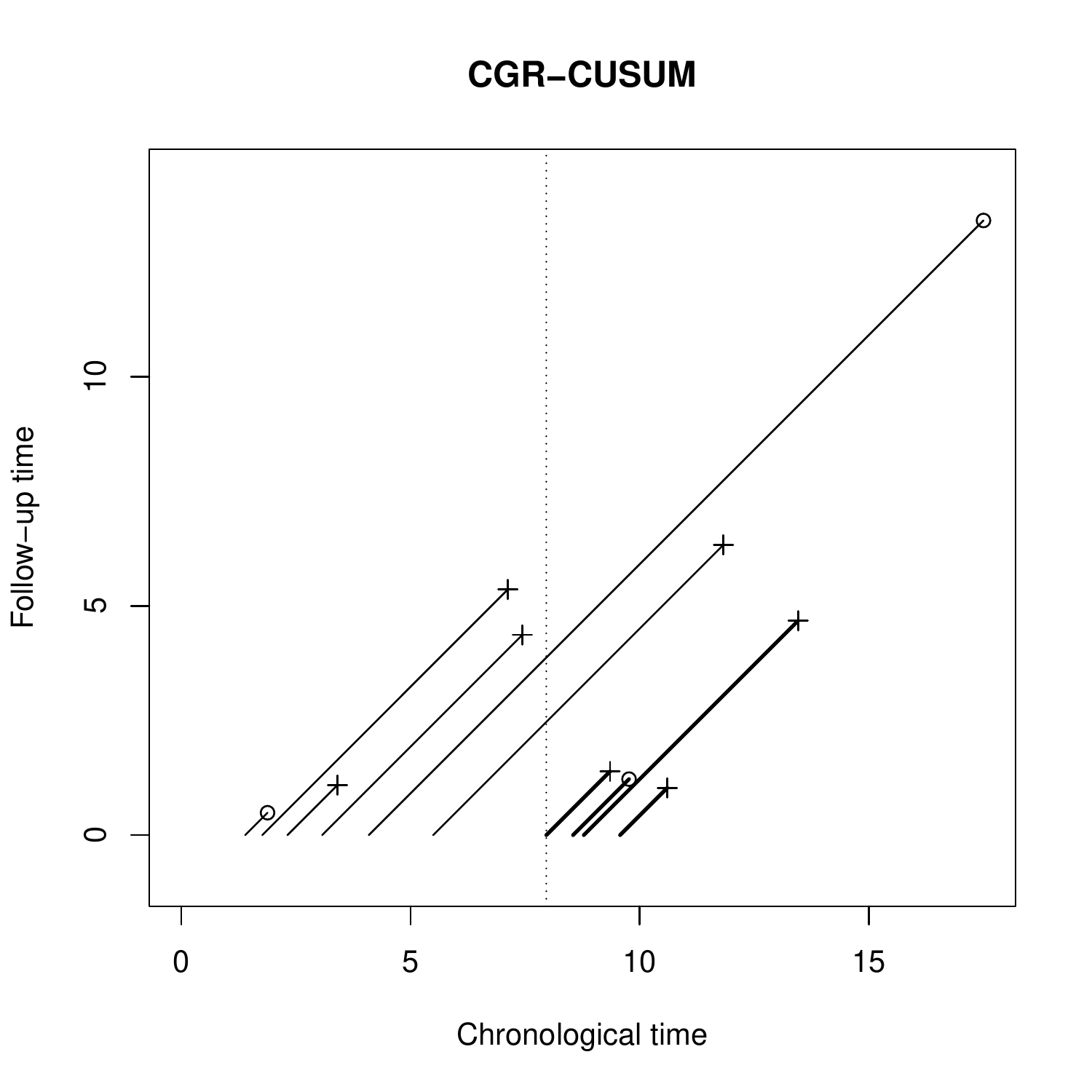}
    \caption{Lexis diagram of CGR-CUSUM}
    \end{subfigure}
    \caption{Lexis diagrams of BK-CUSUM and CGR-CUSUM charts. The line segments in bold represent the information used by the chart if both charts were to detect the same change point (dotted vertical line).}
    \label{fig:lexis}
\end{figure}

\begin{table}
\centering
\begin{tabular}{ |l|c|c|c|c| }
\hline
Hospital nr. & \makecell{Funnel plot \\ $p = 0.95$ \\ yearly}  & \makecell{Bernoulli CUSUM \\ $h=3.5, e^\theta = 2$ \\ monthly} &  \makecell{BK-CUSUM \\ $h=5.1, e^\theta = 2$ \\ monthly}& \makecell{CGR-CUSUM \\ $h=6.8, e^{\hat{\theta}(t)} \leq 6$ \\ monthly} \\\hline
5 & 36 & 31 & \cellcolor{red} & \cellcolor{red} \\ \hline
  9 & 36 & 30 & 21 & 19 \\ \hline
  13 & 36 & 15 & 20 & 21 \\ \hline
  17 & 36 & 24 & 15 & 12 \\ \hline
  19 & 36 & 31 & \cellcolor{red} & \cellcolor{red} \\ \hline
  22 & 36 & 16 & 8 & 7 \\ \hline
  23 & 36 & 29 & 25 & 25 \\ \hline
  32 & 36 & 23 & 24 & 27 \\ \hline
  37 & 36 & 27 & 21 & 18 \\ \hline
  46 & 36 & 30 & 25 & 21 \\ \hline
  48 & 36 & 25 & 18 & 19 \\ \hline
  74 & 36 & 32 & 24 & 19 \\ \hline
  80 & 36 & 27 & 19 & 18 \\ \hline
  11 & 48 & \cellcolor{red} & 24 & \cellcolor{red} \\ \hline
  39 & 48 & 40 & 40 & 42 \\ \hline
  42 & 48 & \cellcolor{red} & \cellcolor{red} & \cellcolor{red} \\ \hline
  58 & 48 & \cellcolor{red} & 40 & \cellcolor{red} \\ \hline
  87 & 48 & \cellcolor{red} & \cellcolor{red} & \cellcolor{red} \\ \hline
  63 & 60 & 58 & 55 & \cellcolor{red} \\ \hline
  81 & 60 & 52 & 59 & \cellcolor{red} \\ \hline
  2 & 72 & 56 & 48 & 48 \\ \hline
  8 & 72 & 63 & 39 & 59 \\ \hline
  73 & 72 & \cellcolor{red} & \cellcolor{red} & \cellcolor{red} \\ \hline
  4 & \cellcolor{red} & 64 & 54 & \cellcolor{red} \\ \hline
  6 & \cellcolor{red} & 52 & 40 & 46 \\ \hline
  18 & \cellcolor{red} & 60 & 54 & 53 \\ \hline
  26 & \cellcolor{red} & 44 & \cellcolor{red} & \cellcolor{red} \\ \hline
  29 & \cellcolor{red} & 48 & 39 & 20 \\ \hline
  35 & \cellcolor{red} & 70 & 43 & 64 \\ \hline
  41 & \cellcolor{red} & 43 & \cellcolor{red} & 33 \\ \hline
  44 & \cellcolor{red} & \cellcolor{red} & 71 & \cellcolor{red} \\ \hline
  50 & \cellcolor{red} & 62 & \cellcolor{red} & \cellcolor{red} \\ \hline
  55 & \cellcolor{red} & 61 & 59 & 51 \\ \hline
  60 & \cellcolor{red} & 66 & \cellcolor{red} & \cellcolor{red} \\ \hline
  68 & \cellcolor{red} & \cellcolor{red} & 59 & \cellcolor{red} \\ \hline
  83 & \cellcolor{red} & \cellcolor{red} & \cellcolor{red} & 39 \\ \hline
\end{tabular} 
\caption{Detection speed of charts in months on the LROI data set. Red cells indicate that this method did not yield a detection on the corresponding hospital before $01/01/2020$. Rows are sorted first according to funnel plot detection time and hospital number afterwards.}
\label{table:detecttimes}
\end{table}

\begin{table}[!ht]
\centering
\begin{tabular}{@{\extracolsep{6pt}} lcc  @{}}
\textbf{N = 97 hospitals} & \textbf{Median (IQR)} & \textbf{Range} \\
\hline
\textbf{Continuous Variables} &&  \\
\cline{1-1}
\hspace{3mm} Mean age (years)&  69.5  (66.8 -  70.2)& 51.8 - 81.5 \\
\hspace{3mm} Mean BMI (kg/m$^2$)& 27.2   (26.9  - 27.5)&  21.2 - 28.4\\
\textbf{Discrete Variables} &&  \\
\cline{1-1}
\hspace{3mm} Gender (\%) & & \\
\hspace{6mm} Female &  65.3  (63.2 -  67.1)& 17.5 - 100 \\
\hspace{6mm} Male& 34.7   (32.9  - 36.8)& 0 - 82.5    \\
\hspace{3mm} Smoking (\%) & &\\
\hspace{6mm} Yes&  11.6  (10 -  13.4)&  0 - 18.4\\
\hspace{6mm} No&  88.4  (86.6 -  90)& 81.6 - 100\\
\hspace{3mm} ASA Classification (\%) & & \\
\hspace{6mm} I&  15.5  (13.1 -  20.4)&  0 - 53\\
\hspace{6mm} II&  63.7  (59.4 -  68.2)& 43.8 - 93.8\\
\hspace{6mm} III-IV&  19.2  (12.6 -  24.6)&  0 - 50\\
\hspace{3mm} Charnley Score (\%) & & \\
\hspace{6mm} A& 47.1   (41.3  - 51.1)& 0 - 76.7\\
\hspace{6mm} B1 &  29  (25 -  33)&  7.1 - 50\\
\hspace{6mm} B2&  21.5  (19.4 -  23.4)&  5.5 - 50\\
\hspace{6mm} C &   2.3  (1.2 -   3.8) & 0 - 16\\
\hspace{3mm} Diagnosis (\%) & & \\
\hspace{6mm} Osteoarthritis&  86.9  (83.7 -  90.3)&  0 - 98.8\\
\hspace{6mm} Not Osteoarthritis&  13.1  (9.7 -  16.3) & 1.2 - 100\\
\textbf{Statistics} &&  \\
\cline{1-1}
\hspace{3mm} Procedures (number) &&\\
\hspace{6mm} in 3 years & 756 (454 - 1227) & 0 - 2523 \\
\hspace{6mm} in 6 years &1638 (1036 - 2462)  &2 - 5093 \\
\hspace{3mm} Revision (\%) & & \\
\hspace{6mm} 1 year&   1.7  (1.1 -  2.3) &0 - 10.4\\
\hspace{6mm} end of follow-up&   2.4  (1.6 -   3.3)&  0 - 13.2\\
\end{tabular} 
\caption{Description of the LROI data set}
\label{table:dataset}
\end{table}
\newpage

\bibliographystyle{unsrtnat}
\bibliography{references}  